\documentclass[aps,twocolumn, pre,superscriptaddress, longbibliography, notitlepage]{revtex4-2}
\usepackage{bm,amsmath,graphicx, mathtools,multirow,gensymb}
\usepackage[colorlinks=true,linkcolor=MidnightBlue,urlcolor=black,citecolor=MidnightBlue,anchorcolor=MidnightBlue]{hyperref}
\usepackage[dvipsnames]{xcolor}
\begin{document}
\title{Minimal mechanism for flocking in phoretically interacting active particles}
\author{Arvin Gopal Subramaniam}
\thanks{Contributed equally}
\email{ph22d800@smail.iitm.ac.in}
\affiliation{Department of Physics, Indian Institute of Technology Madras, Chennai 600036, India}
 \author{Sagarika Adhikary}
\thanks{Contributed equally}
\email{a.sagarika@physics.iitm.ac.in}
\affiliation{Department of Physics, Indian Institute of Technology Madras, Chennai 600036, India}
 \author{Rajesh Singh}
\email{rsingh@physics.iitm.ac.in}
\affiliation{Department of Physics, Indian Institute of Technology Madras, Chennai 600036, India}
\begin{abstract}
Coherent collective motion is a widely observed phenomenon in active matter systems. 
Here, we report a flocking transition mechanism in a system of chemically interacting active colloidal particles sustained purely by chemo-repulsive torques at low to medium densities. The basic requirements to maintain the global polar order are excluded volume repulsions and long-ranged repulsive torques. This mechanism requires that the time scale individual colloids move a unit length to be dominant with respect to the time they deterministically respond to chemical gradients, or equivalently, pair colloids sliding together a minimal unit length before deterministically rotating away from each other. Switching on the translational repulsive forces renders the flock a crystalline structure. 
Furthermore, liquid flocks are observed for a range of chemo-attractive inter-particle forces. Various properties of these two distinct flocking phases are contrasted and discussed. We complement these results with stability analysis of a hydrodynamic model, which admits the transition corresponding to destabilization of the flocking state observed in particle-based simulations. 
\end{abstract}
\maketitle
\maketitle
\section{Introduction}\label{sec:intro}
The collective swarming behaviour of interacting agents - also called flocking \cite{toner2024physics} -  has been a topic of sustained interest in the field of non-equilibrium statistical physics. Flocking is also a ubiquitous phenomena in the natural world, widely reported in groups of animals,
such as birds, fish, and mammals \cite{couzin2002collective, marchetti2013hydrodynamics, vicsek2012collective}. 
Paradigmatic models that incorporate local alignment interactions to study flocking have been extensively studied
\cite{vicsek1995novel,chate2008collective, mishra2010fluctuations, toner2005hydrodynamics, toner2024physics} and continue to be investigated \cite{toner2024_BD,ikeda2024minimum, chate2024dynamic, jentsch2024new}. 
Most notably, in the Vicsek model \cite{vicsek1995novel}, local alignment interactions among the individual agents can lead to transition from a disordered state to one with large-scale, coordinated movement. 
The flocking transition in these models with short-range alignment interactions has been widely studied, \textcolor{black}{along with extensive investigation on the nature of the order-to-disorder transition  \cite{chate2008collective, toner2005hydrodynamics, adhikary2022pattern, adhikary2023collective} and universality of critical exponents \cite{toner2024physics}.}
\\
 
 The possibility of attaining an emergent global polar order (flocking) arising purely out of dynamical particle-based models, without any explicit alignment
interaction, is a subject of recent interest, and recently \textcolor{black}{reviewed in \cite{baconnier2025self}. Systems where such a phase has been observed are typically over-damped \cite{chen2023molecular, grossmann2020particle, hiraiwa2020dynamic} and include those with velocity alignment interactions \cite{grossmann2020particle, chen2024emergent, martin2018collective, sese2018velocity}, distance-dependent interactions \cite{knevzevic2022collective}, history-dependence \cite{kumar2023emergent, subramaniam2024rigid}, and those with repulsive torques and forces between the particles without volume exclusion \cite{bricard2013emergence, das2024flocking}. A separate class of inertial models have also reported such a transition \cite{grossman2008emergence, hanke2013understanding, lam2015self, miranda2025collective}, including those with attractive interactions \cite{caprini2023flocking}.}\\
 
Though various possible mechanisms deviating from the Vicsek-like rule have been established, the explicit relevance of the long-ranged interparticle forces and torques between active particles have not been established. For instance, migrating cells that display a global polarity have been known to display chemo-tactic attractive interactions in addition to (among others) either a contact ``inhibited" or ``attracted" locomotion - akin to a long-ranged torque of either repulsive or attractive nature \cite{camley2017physical, hayakawa2020polar, hiraiwa2020dynamic}. 
\textcolor{black}{Studies on translational and rotational diffusiophoretic motion of active particles (with self-generated chemical field), with attractive torques results collapse of particles into dense clusters \cite{pohlStarkPRL2014}. However, extensive study on repulsive torque is rare in these systems. Thus, open questions remain on the minimal ingredients and mechanism at play arising from the forces and torques of inter-particle 
phoretic interactions.} \\

In this paper, we report a minimal mechanism for the formation of liquid flocks, 
\textcolor{black}{which we herein denote as ``Chemorepulsive Liquid
Flocks" (CLF).}
The basic ingredients that are \textit{sufficient} to produce such a flock are: (i) short-ranged excluded volume repulsion, and (ii) long-ranged (chemo) repulsive torques. There is no need for any additional (chemo)repulsive translational forces between the particles, nor is there a requisite role for noise in destabilizing the flock. 
Furthermore, the incorporation of long-ranged translational chemo-repulsive forces between the particles induces instead a crystalline flock (with regular lattice spacing), 
\textcolor{black}{we denote this herein as ``Chemorepulsive Crystalline
Flocks" (CCF).} Our work thus adds insights to investigations on attaining a global polar order via dynamical mechanisms without any explicit alignment interaction, by providing a minimal mechanism, both for flock formation and destabilization, that, to our best knowledge, has not been reported elsewhere.\\

The remainder of the paper is organized as follows. 
In Section \ref{sec:Model}, we describe our model to study phoretic (chemical) interactions of active particles, \textcolor{black}{and enumerate various important dimensionless parameters of our study.
In Section \ref{sec:CLF}, we report various simulation results for the CLF, and likewise for the CCF in section \ref{sec:CCF}. 
A comparison of these two phases is done in section \ref{sec:CCF_CLF} by studying the pair correlation, hexatic order, order-to-disorder transition, density fluctuations and finite-size effects.
In Section \ref{sec:chi_t_pos}, we demonstrate that flocking persists even for chemo-attractive forces between particles which turn away from each other. 
Further, we complement the results from particle-based simulation with a stability analysis of a continuum model of coupled density and polarization fields in Section \ref{sec:continuum}.
Finally, conclusions and a discussion of our work in relation to other findings is presented in Section \ref{sec:conc}.} \\



 \section{Model}\label{sec:Model}
\textcolor{black}{The model consists of two dynamical parts: (a) the dynamics of particles, and (b) the dynamics of the chemical concentration field; these are described in the following subsections. }\\

\subsection{Dynamics of the particles}
We consider a set of $N$ chemically interacting active colloids of radius $b$. We model the $i$th active particle as a colloid particle centered at $\mathbf r_i=(x_i,y_i)$, confined to move in two-dimensions, which self-propels with a speed $v_{0}$,  along the directions $\mathbf e_i =(\cos\theta_i,\,\sin\theta_i)$. Here $i=1,2,3,\dots,N$ and $\theta_i$ is the angle made by the orientation $\mathbf e_i$ with respect to the positive $x$-axis.
The orientation $\mathbf e_i$ of the $i$th particle, given in terms of the angle $\theta_i$, changes due to coupling its dynamics to the dynamics of a phoretic (scalar) field $c$, as we describe below. The position $\mathbf r_i$ and orientation $\mathbf e_i$ of the $i$th particle is updated as:
\begin{subequations}
\begin{align}
     \dot{\mathbf r}_i  &=  v_{0} \mathbf e_i +\chi_t\, \mathbf J_i + \mu \mathbf F_i + \bm \eta^t_i,\\
     \qquad\dot{ \mathbf e}_i&=\chi_r 
     \Big[
    \left(\mathbf e_i\times\mathbf {\mathbf{{J}}}_i 
    \right) + \bm \eta^r_i \Big] \times \mathbf e_i.
\end{align}
\label{eq:mainLE}
\end{subequations}
Here, $v_0$ is the self-propulsion speed of an isolated active particle. 
\textcolor{black}{The terms $\bm \eta^t_i$ and
$\bm \eta^r_i$
are white noises with zero mean and no temporal correlation. 
The variance of noises $\bm \eta^t_i$ and
$\bm \eta^r_i$, are respectively,
$2D_t$ and $2D_r$. 
These noises terms are included for the sake of completion and to demonstrate that the results are robust against weak fluctuations.} The term $\mathbf J_i$ is chemical flux on the location of the $i$th particle, which gives the inter-particle interaction (of chemical origin). The phoretic flux is given as: $\mathbf {\mathbf{\mathbf{{J}}}}_i(t) =-\left[\mathbf\nabla c(\mathbf r, t)\right]_{\mathbf r=\mathbf r_i}$, 
where $c(\mathbf r, t)$ is the concentration of the chemicals 
(e.g filled micelles in an oil-emulsion system \cite{hokmabad2022chemotactic, kumar2023emergent, dwivedi2022self}).\\

\begin{figure}
    \includegraphics[width=0.48\textwidth]{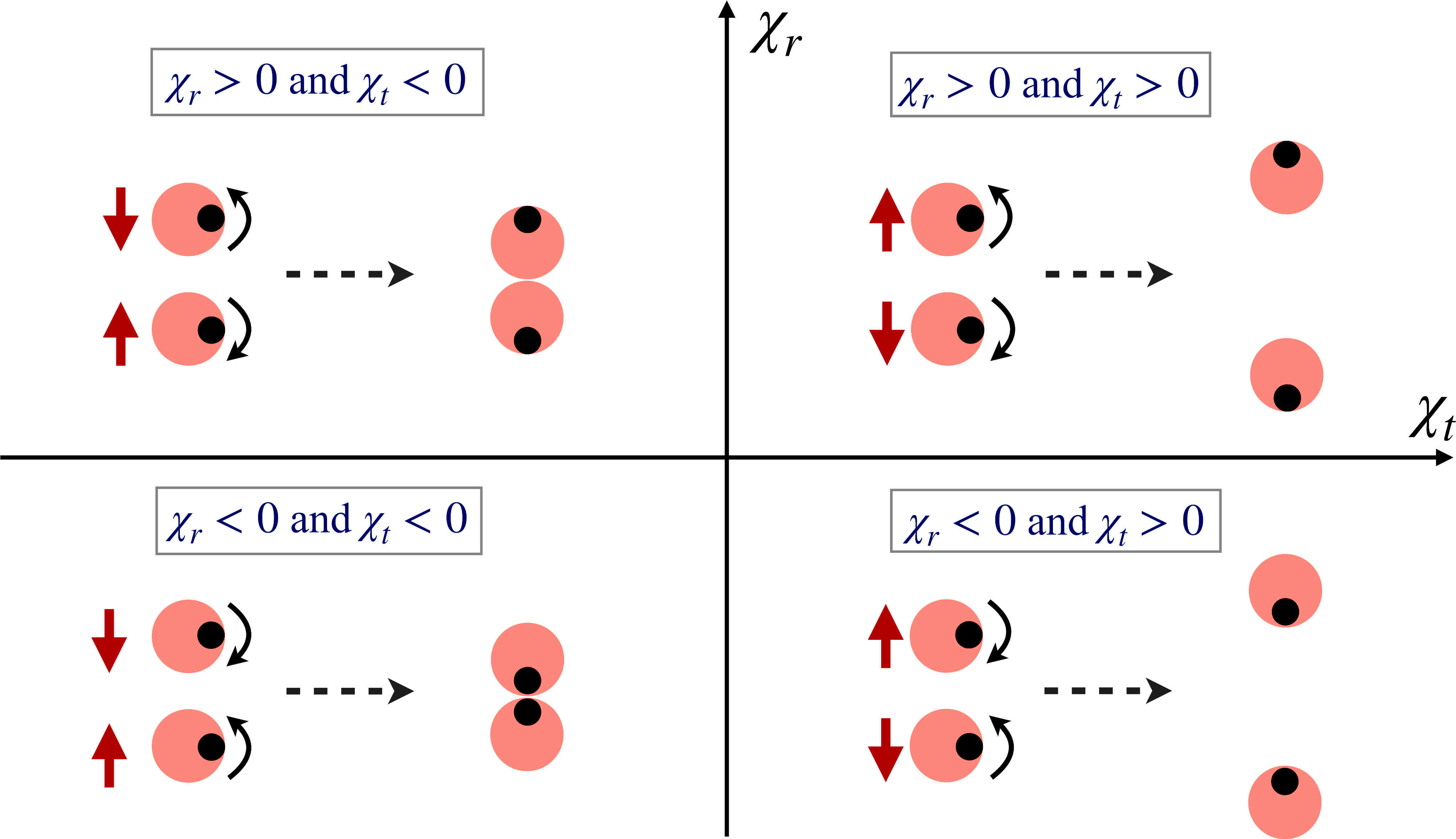}
    \caption{\textcolor{black}{Schematic description of the chemical interactions between the particles based on Eq.\eqref{eq:mainLE}. The black dot on the orange circles indicate the
orientation of the particles. Black curved arrows show rotations from chemical interactions, while solid red arrows shows translations  from chemical interactions. The particles \emph{turn away} from each other if $\chi_r>0$, while they \emph{turn towards} each other if $\chi_r<0$.
Similarly, particles \emph{translate away} from each other if $\chi_t>0$,
while they \emph{translate towards} each other if $\chi_t<0$.   }
}
    \label{fig:0_sch}
\end{figure}

The term proportional to 
$\chi_r$ in Eq.(\ref{eq:mainLE}) drives  orientational changes (turning particles away from each other if $\chi_r>0$) through interparticle chemical interactions. In this paper, we only consider case of  $\chi_r>0$ when particles turn away from each other due to chemical interactions \cite{subramaniam2024rigid,kumar2023emergent}.
Similarly, the term proportional to $\chi_t$ ensures repulsion in the positional dynamics if $\chi_t>0$. If both $\chi_r$ and $\chi_t$ are positive, then the system is said to be \textit{chemo-repulsive}. 
\textcolor{black}{A schematic representation of the dynamics due to the model is given in Fig.\ref{fig:0_sch} to described the physical meaning of parameters $\chi_r$ (which controls rotation) and $\chi_t$ (which controls translation) for chemical interactions between the particles.}
\\

In Eq.\eqref{eq:mainLE}, to preclude overlap of the particles, there is a short-ranged purely repulsive body force $\mathbf F_i$ on the particles. 
The expression of the body force on the $i$th particle is given as: 
\begin{align}
    \mathbf F_i = -\frac{\partial U}{\partial \mathbf r_i}, \qquad 
 U= \sum_{i<j}\mathcal U^e   (\mathbf r_i,\mathbf r_{j}).
  \label{eq:bodyForce}
\end{align}
With $b$ as the colloidal radius and $r_{ij}=|\mathbf {r}_i - \mathbf {r}_j|$, we choose $\mathcal U^e$ to be of the form:
$$
\mathcal U^e=\begin{cases}
    \kappa\left(r_{ij}-2b\right)^2,\qquad \quad 
 \text{if } r_{ij}<2b,\\
0,\qquad\qquad\qquad\qquad \text{otherwise.}
\end{cases}
$$
Here,
$\kappa$ is a constant, which determines the strength of the excluded volume repulsion.

\subsection{Dynamics of the chemical field}
\label{sec:chemEvolv}
Though the focus of this paper is on chemical interactions, we note that our analysis is general and maybe applied to other phoretic fields as well.
 The chemical concentration field $c(\mathbf r, t)$ around the particles located at $\mathbf r_i$ is obtained by solving: 
\begin{align}
    D_{c}\nabla^2 c(\mathbf r, t) + \sum_{i=1}^Nc_0\, \delta(\mathbf r- \mathbf r_i) = 0.
\label{eq:cEq}
\end{align}
Here, $D_{c}$ is the diffusion coefficient and $c_0$ is emission constant of the chemical. 
In the above, we have assumed that the chemical interactions between the particles is instantaneous   \cite{subramaniam2024rigid, pohlStarkPRL2014, liebchen2019interactions, saha2014clusters, soto2015self,ruckner2007chemically,singh2019competing, turk2024fluctuating}. In addition, we assume that each particle is a point source of the chemical field. Finally, the chemical flux $\mathbf{{J}}_i$ on the $i$th colloid, 
in the point particle limit, can be written from the solution of the Eq.\eqref{eq:cEq} as:
\begin{align}
    \mathbf{J}_{i} = \frac{c_0}{4\pi D_c}\sum^N_{\substack{j=1\\ i\neq j}} 
 \frac{\mathbf  r_{ij}}{
 r_{ij}^3 }.
 \label{eq:curr_r2}
\end{align}
Here, $r_{ij}=|\mathbf r_{ij}|$, with
$\mathbf r_{ij}=\mathbf{r}_{i} - \mathbf{r}_{j}$. 
\begin{figure*}[t!]
    \includegraphics[width=0.96\textwidth]{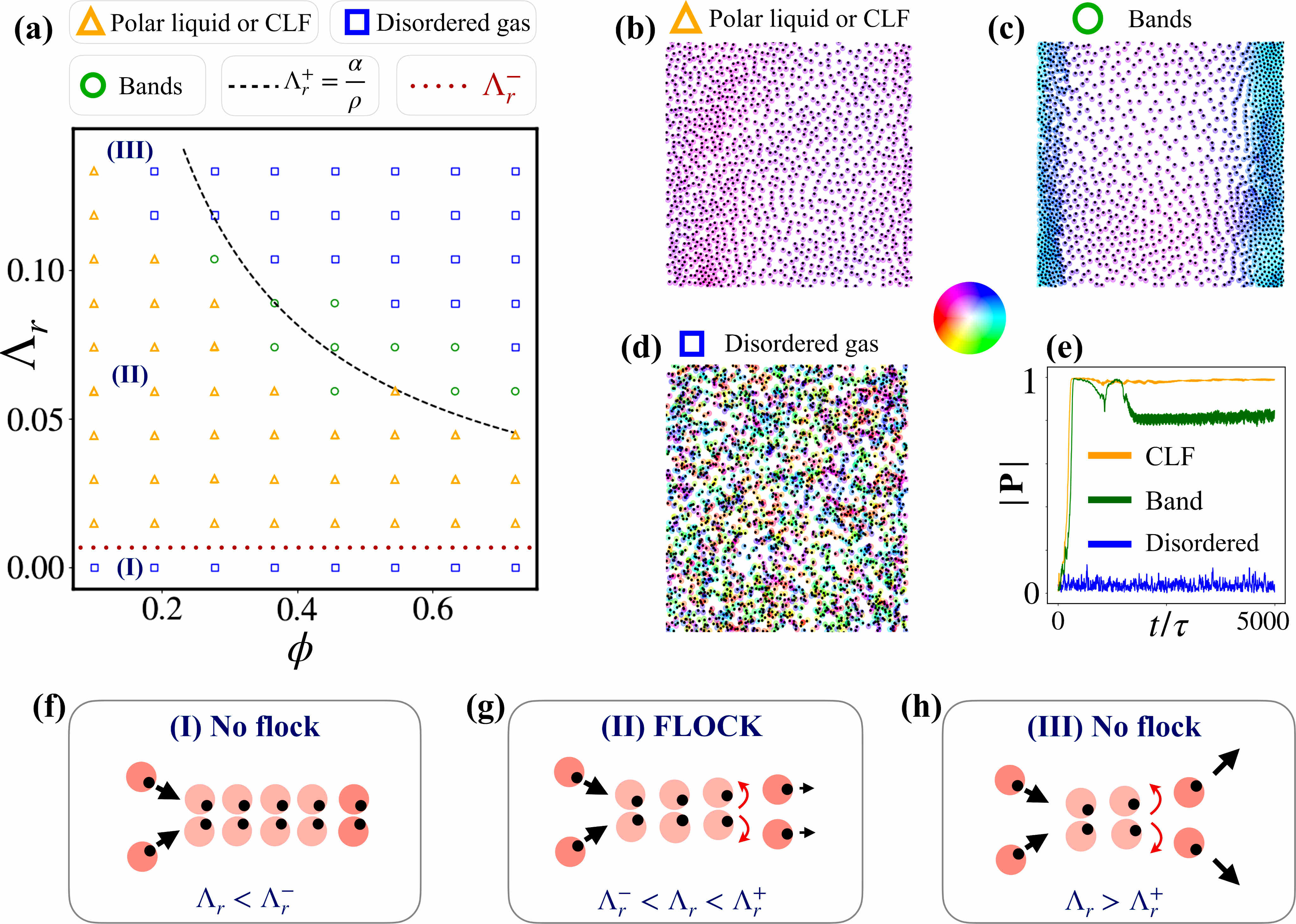}
    \caption{
    \textcolor{black}{CLF for $\Lambda_t=0$.
    (a) Phase diagram in the $(\phi, \Lambda_r)$ plane, where $\phi$ is area fraction of particles. Here, 
    $\Lambda_r^{-}$ is annotated as a dotted red line, while the density dependent $\Lambda_r^{+}$ in dashed black line.
    Symbols here and throughout the paper denoting a flock (yellow triangle), bands (green circles), and disordered (blue squares). Marked (I)-(III) scenario and transition lines (black and red) are described in panels (f-h). Snapshots of the respective phases are shown in (b)-(d) respectively (color coding for orientation of each particle is in middle panel). (e) The evolution of the polarization is displayed for the corresponding states (similarly colored). Panels (f-h) are series of scenarios depicted corresponding to parameters choices labelled in panel (a). Scenarios (I) and (III) have chemical response that are either too slow or too quick, such that a stable flock cannot be formed.
    Panel (g)- scenario (II) - shows the regime in which the chemical response is in an optimal range such the pair collision mechanism that would produce a flock.
    See \ref{sec:threeCases} for details of the three distinct cases and mechanism of flocking.}   
    }
    \label{fig_CLF}
\end{figure*}
It is worthwhile to note that in our model, the chemical field diffuses in an infinite three-dimensional half-space, and thus, the chemical field decays as $1/r$. Assuming a no-flux boundary condition at the bounding surface, the $1/r$ dependence of the chemical field remains valid. On the other hand, the particles are confined to move in two-dimensions, such that they move in the plane of the bounding surface. 
Consequently, the dimensions of chemical concentration is: $[c]=[\text{L}^{-3} ]$, and chemical flux are $[\mathbf J]=[\text{L}^{-4} ]$.
So, the dimensions of parameters are $[\chi_r]=[\text{L}^{4}\text{T}^{-1} ]$ and $[\chi_t]=[\text{L}^{5}\text{T}^{-1} ]$.

\subsection{Important dimensionless parameters}
\textcolor{black}{ We note that $\tau = {b}/{v_{0}}$ is the propulsion time scale. In addition, we define $\tau_r=b^4/\chi_r$, which sets the time scale for deterministic rotation of the particle due to phoretic interactions. 
The phenomenology of the CLF can also be understood by studying the dynamics as per the dimensionless numbers of the problem that arise out of the equations of motion (\ref{eq:mainLE}). They are:
\begin{align}
    \Lambda_{r} &=\frac{\tau}{\tau_r}= \frac{\chi_r }{b^3v_{0}}, \qquad\quad
        \Lambda_{t} = \frac{\chi_t}{b^4 v_{0}}.
\label{eq:dimNumbs}
\end{align} 
Here, $\Lambda_{r}$ is the ratio of propulsion time scale $\tau$ and time scale $\tau_r$ which controls rotations of the particle orientations from chemical interactions between particles. On the other hand, 
$ \Lambda_{t}$ is the ratio of translation due to chemical interactions between particles and one-body propulsion. The key parameters of this model are the dimensionless numbers $\Lambda_r$, $\Lambda_t$ and $\phi$. 
Here, $\phi$ is the  the area fraction. 
$\phi$ is related to the number density $\rho$, as $\phi= {\left(N\pi b^2\right)}/{L^2}=\rho \pi b^2$. The simulation results are henceforth presented as a function of these three parameters. }\\

\textcolor{black}{We note that in principle, there are other relevant dimensionless numbers; e.g. $\text{Pe}_1 = \frac{v_0 b}{D_t}$ and $\text{Pe}_2 = \frac{v_0}{b D_r}$ which would alter the flocking transition. Our results are presented for effectively noiseless systems, thus the aforementioned $\text{Pe}_{1,2} \rightarrow \infty$. Indeed, the novelty of the mechanism presented here, as we will show below, is its purely deterministic destabilization of the polar phase, independent of any noise contributions. }\\


\subsection{Simulation details}
\textcolor{black}{The position and orientations of the particles are updated using a forward Euler-Maruyama method from the dynamical equations given in Eq. \ref{eq:mainLE}. Initially particles
are distributed randomly over the two-dimensional space. The initial orientations are also randomly distributed over
the range of angles [$-\pi,+\pi$] (angles are computed with respect to the positive $x$-axis). Periodic boundary
conditions are applied on both the $x$-axis and $y$-axis. To compute chemical interactions using the expression of
$\mathbf J_i$, we need to sum over periodic images. We have  used a regular summation convention (using minimum image
convention) for all results reported here, and checked these with the full Ewald summation convention for selected
parameter values.  In this paper, the fluctuations (noise terms) are set to be negligibly small to emphasize that flocks can be formed and destroyed by deterministic causes. Indeed, the polar order is expected to be destroyed at larger values of fluctuations \cite{vicsek1995novel}. 
But we do not explore the role of 
the noise in the positional and rotational sectors in this paper. Instead, we choose their strengths to be sub-dominant compared to deterministic effects due to chemical interactions and self-propulsion. We have taken particle radius $b=1$, total number of particles varies from the range $N=[239, 20000]$. The steady-state time averages of the observables are taken over $2 \times 10^5$ time steps. The time-stepping of the Euler-Maruyama integrator $dt$ is taken as $0.01$. 
A summary of all of the values of parameters used for the simulation are given in Table (\ref{si_table_params}) in the Appendices. }\\
%


\section{CLF via chemo-repulsive torques} \label{sec:CLF}

We first consider the case of  with $\Lambda_t=0$ (or, equivalently $\chi_t = 0$). 
Thus, repulsions are exclusively short-ranged via Eq.\eqref{eq:bodyForce} which prevents overlap of particles.
First, we vary $\Lambda_r$ and $\phi$ and determine the polar order of the system. The phase diagram delimiting the CLF phases (polar flocks, bands, and disordered states) are shown in Figure \ref{fig_CLF}(a). The states are delimited via the mean global polarization of the system at the steady state, $P_{\mathrm{ss}}=\langle |\mathbf{P}| \rangle_{\mathrm{ss}}$, where the polarization $\mathbf{P}$ is defined as the average of sum of the orientation of all the particles:
\begin{align}
    \mathbf{P} =\frac1N \sum_{i=1}^N \mathbf{e}_i.
\end{align} 
The different phases shown in the phase diagram are distinguished with different values of the magnitude of $P_{\mathrm{ss}}$: flock ($P_{\mathrm{ss}}>0.9$), bands ($0.1<P_{\mathrm{ss}}<0.9$) and disordered ($P_{\mathrm{ss}}<0.1$). The snapshots of these three phases are presented in Fig. \ref{fig_CLF}(b)-(d) respectively. The dynamics of $| \mathbf{P}|$ of the respective cases is shown in \ref{fig_CLF}(e). On the $(\phi,\Lambda_r)$ plane (Fig. \ref{fig_CLF}(a)), we see that low-density flocks are sustained, that are eventually destabilized at sufficiently high repulsion and densities. These flocks have evidently a liquid-like spatial structure (Fig. \ref{fig_CLF}(b)) \cite{bricard2013emergence, geyer2018sounds, geyer2019freezing}. The density field $\rho$ is thus not homogeneous across the system (see Fig. \ref{fig_CLF}(b)). Density bands often occur in these systems, where particles move perpendicular to the length of these bands and travel Fig. \ref{fig_CLF}(c)). For larger systems, many of these dense bands are observed over the space.
See Movie I \cite{siText} for dynamics to obtain CLF starting with a disordered state. \\

\subsection{Mechanism for the formation of CLF}\label{sec:threeCases}
\textcolor{black}{The CLF flock forming region on Fig. \ref{fig_CLF}(a) can be understood by consideration of competing effects between various length (time) scales of the problem. To this end, it is useful to consider the idealization of pair colloidal collision, where the pair collides move along a co-moving frame a distance $l_{sl}$ within a time $\tau_{sl}$, whilst rotating away (chemorepulsive). }This can be interpreted as a \textit{sliding} length and time scales respectively, and can be written as:
\begin{align}
    \tau_{sl} = \frac{b }{v_{0} \Lambda_r},
    \qquad l_{sl} = v_0 \tau_{sl} = \frac{b}{\Lambda_r}.
    \label{eq:slidingTL}
\end{align}


Let us denote $\Lambda_r^{-}$ as the smallest value of $\Lambda_r $ to sustain the CLF (corresponding to the lowest transition line in Fig.\ref{fig_CLF}a), and  $\Lambda_r^{+}$ as the largest value of $\Lambda_r $ above which the CLF is not sustained. $\Lambda_r^{+}$ is density dependent, as will be described below. The mechanism for the CLF can then be understood as a competition between $\Lambda_r$ and $\Lambda_r^{+}$, $\Lambda_r^{-}$; this is illustrated in Fig. \ref{fig_CLF} (f-h) (annotated accordingly in the phase diagram of Fig.\ref{fig_CLF}(a)). 
From (\ref{eq:slidingTL}), the mechanism can be understood in terms of sliding time scale
$\tau_{sl}$ along with $\tau_{sl}^-$ and $\tau_{sl}^+$.
There are three distinct cases (see corresponding label on phase diagram in Fig. \ref{fig_CLF}(a)):\\

\textbf{Case I.}
When $\Lambda_r  < \Lambda_r^{-}$ 
(corresponding to $\tau_{sl} > \tau_{sl}^{-} $), the orientational changes of the colloids are too slow compared to the free movement time, pair colloids slide excessively long, hence local polar order cannot be sustained (Fig.~\ref{fig_CLF} (f)). \\

\textbf{Case II.} In the regime of $ \Lambda_r^{-} < \Lambda_r  < \Lambda_r^{+}$ (equivalently $\tau_{sl}^{+} < \tau_{sl} < \tau_{sl}^{-}$), we have that the chemical response time fast enough such that local polar order is sustained, but still sufficiently slow such that the forward-backward symmetry at the pair collision level is broken (Fig.~\ref{fig_CLF} (g)). 
 \\

\textbf{Case III.}
Finally, for $\Lambda_r  > \Lambda_r^{+}$ (equivalently $\tau_{sl} < \tau^+_{sl}$), the chemical response is sufficiently quick such that there is no symmetry breaking at the pair collision level, and hence no local alignment (See Fig.~\ref{fig_CLF} (h)). \\

\subsection{Flocking by sliding an optimal distance while turning away}
Using Eq.\eqref{eq:slidingTL}, we may also describe the mechanism for CLF in terms of sliding length scale $l_{sl}$, corresponding to 
$\tau^+_{sl}$ and $\tau^-_{sl}$. These are: 
$$l_{sl}^{+}=v_0 \tau^+_{sl},\qquad l_{sl}^{-}=v_0 \tau^-_{sl}.
$$
\textcolor{black}{Note that in this choice of notations, $l_{sl}^{-} > l_{sl}^{+}$ as the time scale $\tau^-_{sl}>\tau^+_{sl}$.
From simulations at $\phi \approx 0.31$, we find these values to be $l_{sl}^{+} \sim 8.3b$ ($\Lambda_r^{+} \sim 0.12$) and $l_{sl}^{-} \sim 100b$ ($\Lambda_r^{-} \sim 0.01$) , thus for low densities $l_{sl}^{+}$ of a few times the colloidal radius is feasible, whereas the upper limit spans the box size, which prohibits any flock formation. As we will show (and derive) below, $l_{sl}^{+}$ is directly proportional to density (with $l_{sl}^{-}$ fixed by the system size), thus CLF formation can be equivalently understood by the competing effects of density-dependent inter-particle (pair) collisions and sliding. 
We see that the phase boundary in Fig. \ref{fig_CLF}(a) takes the functional form $\Lambda_r^{+}  = {\alpha}/{\rho}$. Here, $\alpha$ is constant that we will later quantify and derive. 
For densities above a critical density $\rho_c = (\frac{\alpha}{b})^{2/3} \approx 0.05$, $l_{sl}^{+}$ always exceeds the mean free path $\frac{1}{\sqrt{\rho}}$ so that flock formation can happen uninhibited. Indeed for very dilute systems ($\rho < 0.05$, smaller than the densities displayed in Fig. \ref{fig_CLF}), there is no flock formation as there is a large free path movement and thus pair sliding is infrequent (not shown here). For high densities, though pair collisions are more frequent, $l_{sl} > l_{sl}^{+}$ cannot be satisfied. Indeed, at these densities, $l_{sl}^{+}$ and $l_{sl}^{-}$ converge and flock formation is prohibited. Thus, for a finite range of densities the CLF can form and stably persist. Equivalently, there is no CLF if the pair colloids slide is too short ($l_{sl}<l_{sl}^+$) or too long ($l_{sl}>l_{sl}^-$).}\\

\textcolor{black}{We clarify at this juncture that such a pair collision mechanism is constructed as an idealization; in actual simulation, pair colloids may not slide an entire $l_{sl}$ before leaving a given collision radius. However, such an abstraction helps us understand the polar alignment mechanism at the pair colloidal level, and in addition agrees fully with simulation results at the collective level. A key feature of this mechanism to be emphasized is the requirement of short-ranged repulsion (provided here via excluded volume interactions), without which CLF formation at $\chi_t=0$ would not occur; this instead would have to be supplemented with $\chi_t>0$ \cite{das2024flocking}. Thus, (i) short-ranged translational repulsion and (ii) long-range orientational torques are the \textit{minimal requisite} ingredients required for the formation of CLF.} \\

\begin{figure}
    \centering
    \includegraphics[width=0.33\textwidth]{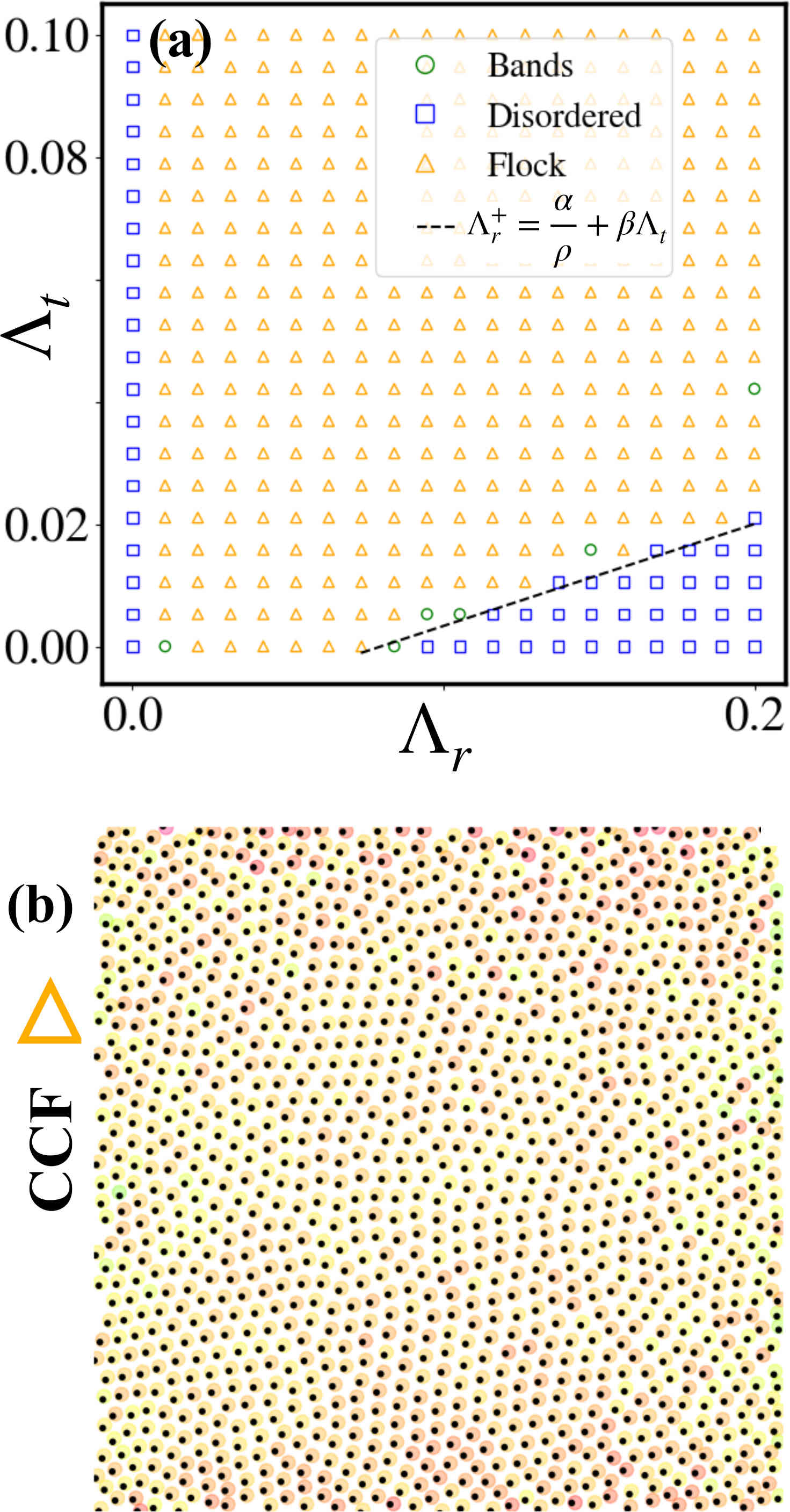}
    \caption{ 
    (a) Phase diagram in the $(\Lambda_r, \Lambda_t)$ plane for the case of $\chi_t > 0$. Dashed line refers to lower bound on $\chi_r$, solid line for lower bound on $\chi_r$ from Eq. (\ref{eq:ineq_chit_chir}). (b) 
    shows a snapshot of the CCF phase (same color scheme to indicate the orientation of the particles as in Fig (\ref{fig_CLF})). 
    }
\label{fig_CCF}
\end{figure}

\section{CCF via chemo-repulsive forces and torques}
\label{sec:CCF}
We now switch on $\chi_t$ and ask how the properties of the flock differ. 
On the $(\Lambda_r,\Lambda_t)$ plane, we find that there is a clear region of flocking for non-zero $\chi_t$ (see Figure \ref{fig_CCF}(a)). We further find that the flocks have crystalline order (see Figure \ref{fig_CCF}(b)). 
 Such crystalline flocks in repulsive systems have been recently reported \cite{das2024flocking}. Similarly to the CLF, the CCF destabilized by sufficiently high rotational torques. In addition, $\Lambda_r^{+}$ now is linear in $\Lambda_t$. $\Lambda_r^{+}$ being larger for a positive $\Lambda_t$, this thus renders a shorter sliding length for pair colloids during the CCF formation. The net result is thus a shorter sliding length compared to the CLF, which we can attribute to the difference between crystalline and liquid structures. See Movie II \cite{siText} for dynamics to obtain CCF starting with a disordered state. 
 We note that if this repulsion is too strong, the CCF is destabilized to a disordered gas (not shown in Figure \ref{fig_CCF}). \\
 
The phase diagrams for the collective dynamics of the system have been presented in Figures (\ref{fig_CLF})(a) and (\ref{fig_CCF})(a) in terms of the three key dimensionless parameters: $\Lambda_t$, $\Lambda_r$ and $\phi$.
 The most generic form of the transition lines from the phase diagrams can be written as:
\begin{align}
    \Lambda_r< \Lambda_r^{+},\qquad \Lambda_r^{+}= \frac{\alpha}{\rho} + {\beta\Lambda_t}.
\label{eq:ineq_chit_chir}
\end{align}
 We obtain $\beta$ and $\alpha$ by first fitting the boundary line in Fig. \ref{fig_CCF}(a). The slope gives $\beta \approx 0.167$, whilst the $x$-intercept gives $\alpha \approx 0.013/b^3$. This fit is obtained by first applying determining points of greatest variance of the order parameter (susceptibility) on the phase plane, and then applying least-squares regression on those points. The fitted $\alpha$ thus independently fixes the boundary on Fig. \ref{fig_CLF}(a).  
The term  $\alpha/\rho$ inhibits sliding at high densities. 
The second term due to the presence of long-ranged translational repulsion suppresses the sliding length and contributes to the crystalline structure. The delimiting line of $\Lambda_r^{-}$ on the left (vertical) of Fig. \ref{fig_CCF}(a) and bottom (horizontal) of Fig. \ref{fig_CLF}(a) are equal. Thus, the CCF mechanism may be summarized by being an extension of the CLF mechanism, with the additional long-ranged repulsive forces between the particles rendering the crystalline structure. \\ 

\section{Differences between CCF and CLF}
\label{sec:CCF_CLF}

\begin{figure}[b!]
    \centering
    \includegraphics[width=0.45\textwidth]{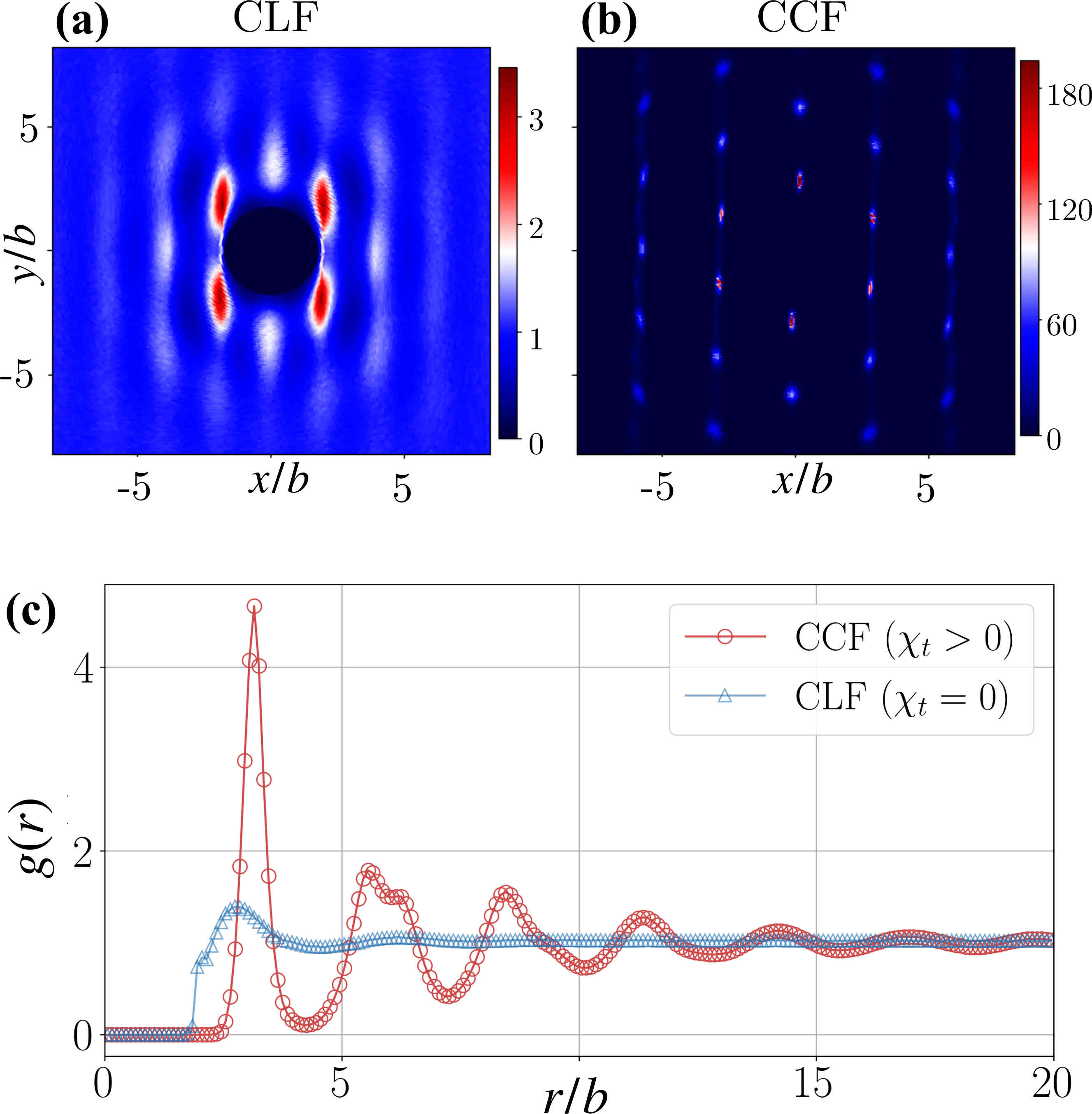}
    \caption{ 
 (a) and (b) compare $g(r, \varphi)$ for the case of the CLF and the CCF. In (c), a comparison of the flocking phase, for CLF (red) and CCF (blue), is shown for the pair correlation $g(r)$. 
    }
\label{fig:CCF-CLF_gr}
\end{figure}
\subsection{Pair correlation function} 
The difference in spatial structure between the CLF and CCF is readily distinguished by the pair correlation function, which is given by:
\begin{align}
    g(r) = \frac{1}{N} \sum_{i,j} \langle \delta \left( |\mathbf{r}_{i} - \mathbf{r}_{j}| - r \right) \rangle
\label{eq:pair_corr}
\end{align}
This is plotted in Fig. (\ref{fig:CCF-CLF_gr})(c), distinguishing the solid and liquid signatures \cite{bricard2013emergence}. A closely related quantity is the polar pair correlation $g(r, \varphi)$ which explicitly depends on the orientations of the colloids. Here, $\cos (\varphi_{ij}) = \mathbf{e}_{i} \cdot \hat{\mathbf r}_{ij}$, where $\hat{\mathbf r}_{ij}=(\mathbf r_{j} -  {\mathbf r}_{j})/| \mathbf r_{j} -  {\mathbf r}_{j}|$\cite{zhang2021active}. For the CLF, we see that there are quadrants where neighbours are more likely to be found (\ref{fig:CCF-CLF_gr}(a)), as opposed to the hexatic order of the CCF (\ref{fig:CCF-CLF_gr}(b)). In the case of CCF, the intensity (proportional to the probability of finding the particle from the reference - see Appendix  \ref{app:CCFCLF}) in the well-localized hexagonal positions of the crystal are substantially higher than in the CLF, which are instead less intense with a greater relative spread. The four-lobe distribution is indicative of a highly anisotropic liquid phase where there is a decreased probability of finding a neighboring particle either at the front or back of the reference particle due to mutual repulsion. \\ 

 \begin{figure}[h!]
     \centering
     \includegraphics[width=0.5\textwidth]{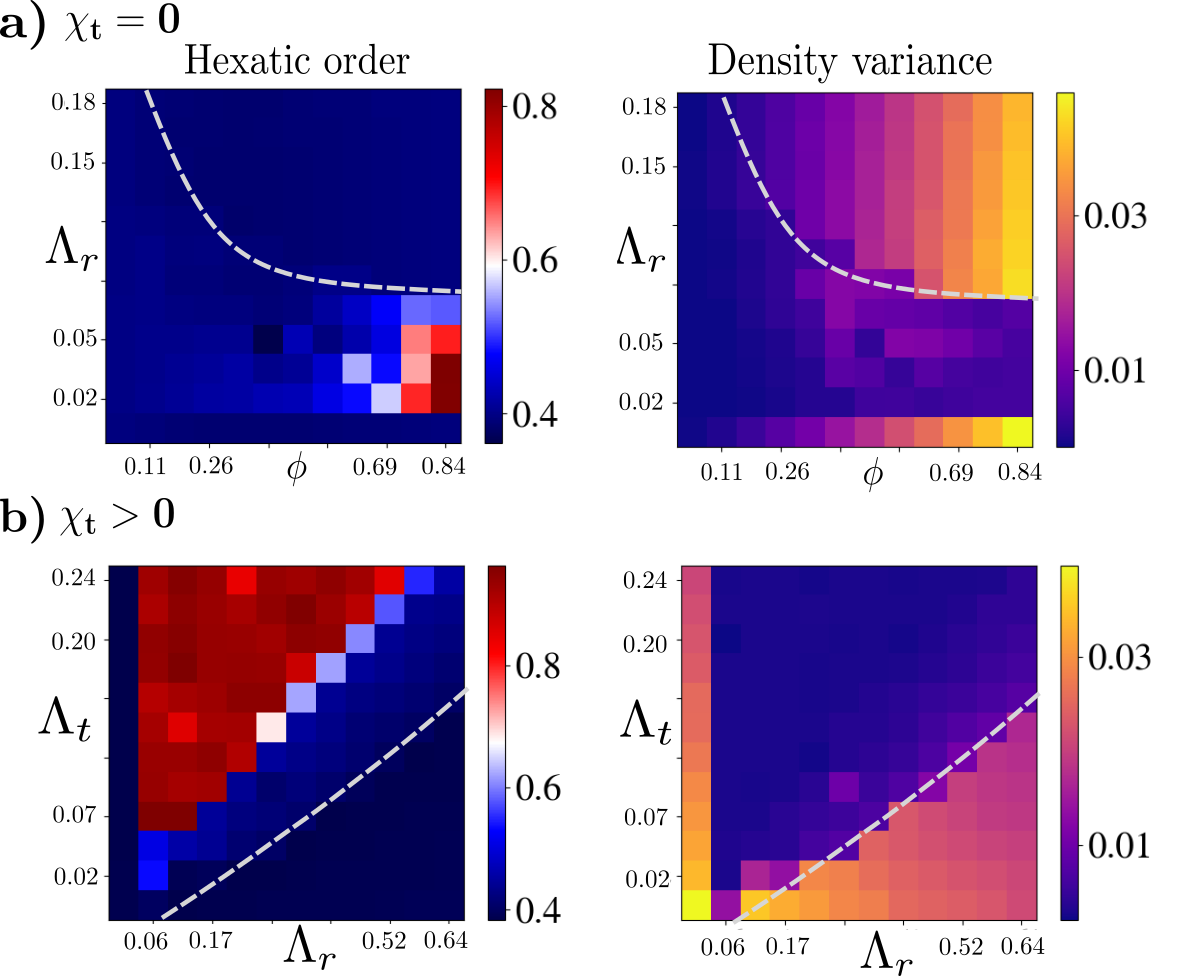}
     \caption{Phase diagrams in terms of (left) global hexatic order ($\psi_6$) and (right) density variance. The \textit{top} panel (a) show these for the CLF, whereas the \textit{bottom} panel (b) for the CCF. Dashed curves in each plot indicate order-disorder (polarization) transition line as in Figures \ref{fig_CLF}(a) and \ref{fig_CCF}(a).}
\label{fig:hex_phase}
 \end{figure}

\subsection{Hexatic order and density variance}
We define a measure of local hexatic order $\psi_i$ for the $i$th particle to quantify the hexatic order globally in terms of $\psi_6$. 
They are given as \cite{chaikin1995principles}:
\begin{align}
    \psi_6= \frac{1}{N} \sum_{i}^{N} \psi_{i},
    \qquad
    \psi_{i} =  \frac{1}{{N^n_i}} \sum_{j}^{{N^n_i}} e^{i6\theta_{ij}}.
\end{align}
Here, $\theta_{ij}$ is the angle between particle $i$ and particle $j$, and ${N^n_i}$ is the number of neighbors for
particle $i$ ($N^n \approx 6$ for all data points). The observable $\psi_6$ is then averaged over $1000$ different
particle configurations at the steady-state. We plot $\psi_6$ for the two cases ($\chi_t=0$ and $\chi_t>0$)
in the left panel of Fig. \ref{fig:hex_phase}. The transition lines from the phase diagrams in Figures \ref{fig_CLF}(a)
and \ref{fig_CCF}(a) are also shown as dotted lines for guide to the eye. In the case of $\chi_t=0$ (Fig. \ref{fig:hex_phase}(a) left), the hexatic order exists in the flocking phase, only for a very high density (area fraction
$\phi=\rho \pi b^2 > 0.6$. Whereas, in case of $\chi_t>0$ (Fig. \ref{fig:hex_phase}(b) left), with $\phi=\rho \pi b^2 =0.36$, the system becomes
highly hexatic in the flocking phase. However, near the transition line, where density bands are prone
to occur, we find that the hexatic order gradually decreases.\\

Next, density variance for the two cases ($\chi_t=0$ and $\chi_t>0$) 
are shown in the right panel of Fig. \ref{fig:hex_phase}. 
The density variance is calculated 
from the density distribution for a specific parameter set shown
in the phase diagram. For local density measurements, the Voronoi tessellation method has been implemented for a
spatial configuration of particles. We calculate local density $\rho_l={\pi b^2}/{A_l}$, where $A_l$ is local area assigned to each particle. Then we compute the density variance as $\langle (\Delta \rho)^{2} \rangle = \langle \rho_l^{2} \rangle - \langle \rho_l \rangle^{2}$ (spatial average) and finally averaged over $1000$ configurations. In both cases, the density variance is much lower in the flocking state. However, it shows an increase near the transition region. The density variance is high for the high area fraction ($\phi=\rho \pi b^2$) and in the disordered phase, for the case $\chi_t=0$ (Fig. \ref{fig:hex_phase}(a) right panel). For $\chi_t>0$ case,
with low $\chi_t$ values, the variance is more prominent, shown in Fig. \ref{fig:hex_phase}(b) (right panel). \\

\begin{figure}[b!]
    \includegraphics[width=0.45\textwidth]{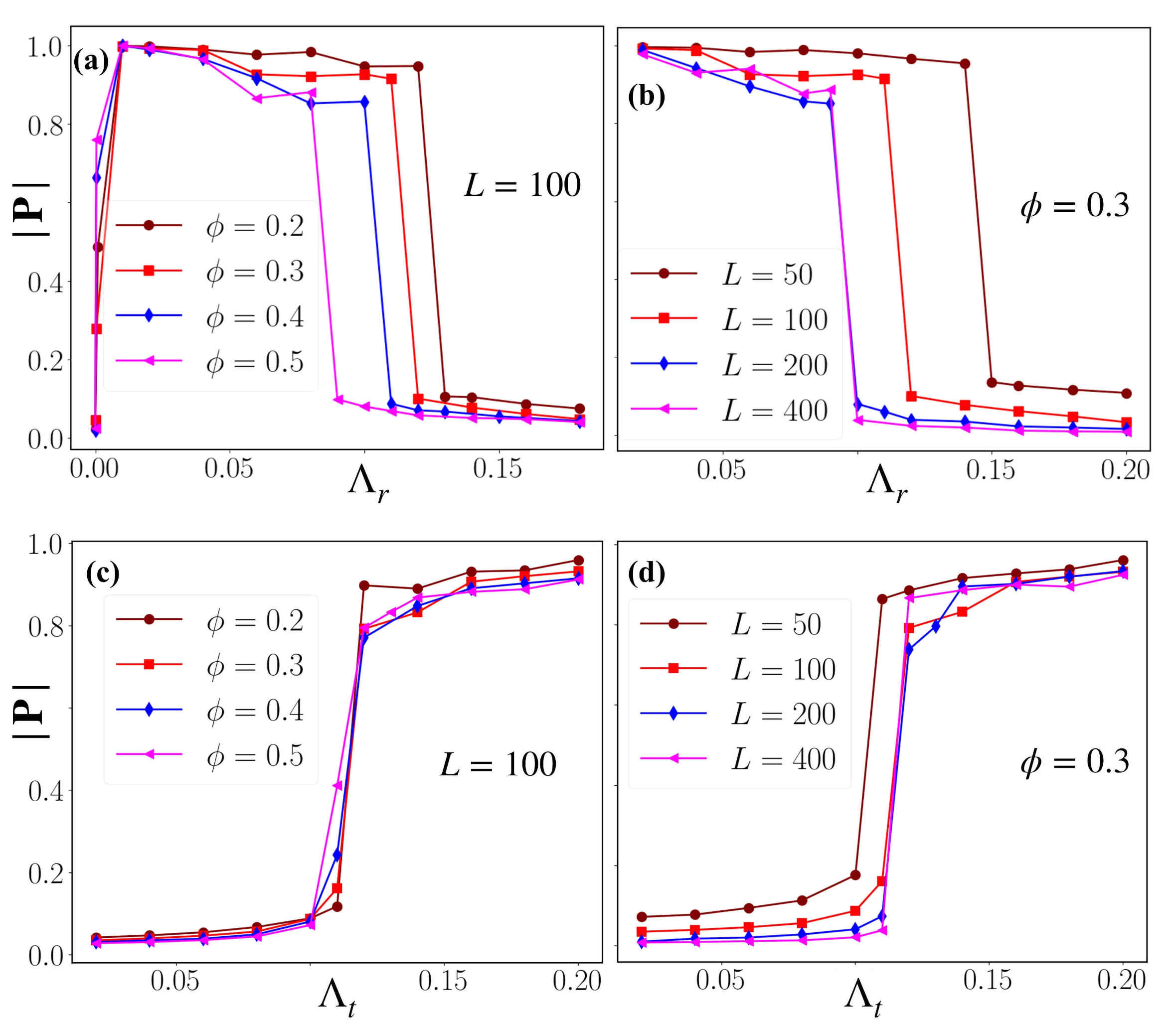}
    \caption{ \textcolor{black}{Finite-size effects of phase transition. \textit{Top} panel : (a) shows magnitude of the polar order parameter $|\mathbf P|$ versus $\Lambda_r$ for the CLF ($\Lambda_t=0$) for different area fractions $\phi$, whilst (b) displays these for different system of sizes $L$. \textit{Bottom} panel: similarly plotted for the CCF, with the variation with respect to $\Lambda_t$ instead for different $\phi$ (c) and $L$ (d) respectively. Here, we have fixed $\Lambda_r =0.55$.  The total number of particles $N$ corresponding to the $L$ are given as: ($L=50, N= 239$), ($L=100, N=955$), ($L=200, N=3820$) and ($L=400, N=15279$).}}
    \label{fig:phase_t}
\end{figure}

\subsection{Phase transition and finite size effects} 
\textcolor{black}{We now present how polar order parameter $|\mathbf{P}|$ changes with variation of the control parameter $\Lambda_r$ and $\Lambda_t$, shown in Fig. \ref{fig:phase_t}. First, for CLF ($\Lambda_t=0$), phase transitions are shown for different area fraction $\phi=[0.2,0.5]$ with system of size $L=100$ (Fig. \ref{fig:phase_t}(a)). With zero $\Lambda_r $, $|\mathbf{P}|\approx0$ and there is no flocking (lower bound of Fig. \ref{fig_CLF}(a)). As we increase $\Lambda_r $, particles flock into a polar ordered state. Whereas with further increase $\Lambda_r $, disordered phase are observed (see Fig. \ref{fig_CLF}(a)). Transition points shift towards lower $\Lambda_r$ values with increase of density. Then we present the finite size effects with different system of sizes $L=[50,400]$ ($N=[239,15279]$) with a constant area fraction $\phi=0.3$ (Fig. \ref{fig:phase_t}(b)). The effects of a finite size on the flocking transition can be clearly seen for smaller systems of sizes, whereas the effect is minor for larger system size. Then, for the CCF (with fixed $\Lambda_r=0.55$), the polar order parameter $|\mathbf{P}|$ versus $\Lambda_t$ is plotted for different area fractions (for reference, see Fig. \ref{fig:hex_phase}(b), where the area fraction is fixed at $\phi=0.36$). In Fig. \ref{fig:phase_t}(d), we study finite-size effects for this case and found the effect on the phase transition is negligible for larger system sizes. It needs to be noted that our model includes long-range interactions, and it is computationally expensive to work with such a system.}

\begin{figure}[b!]
    \centering
    \includegraphics[width=0.475\textwidth]{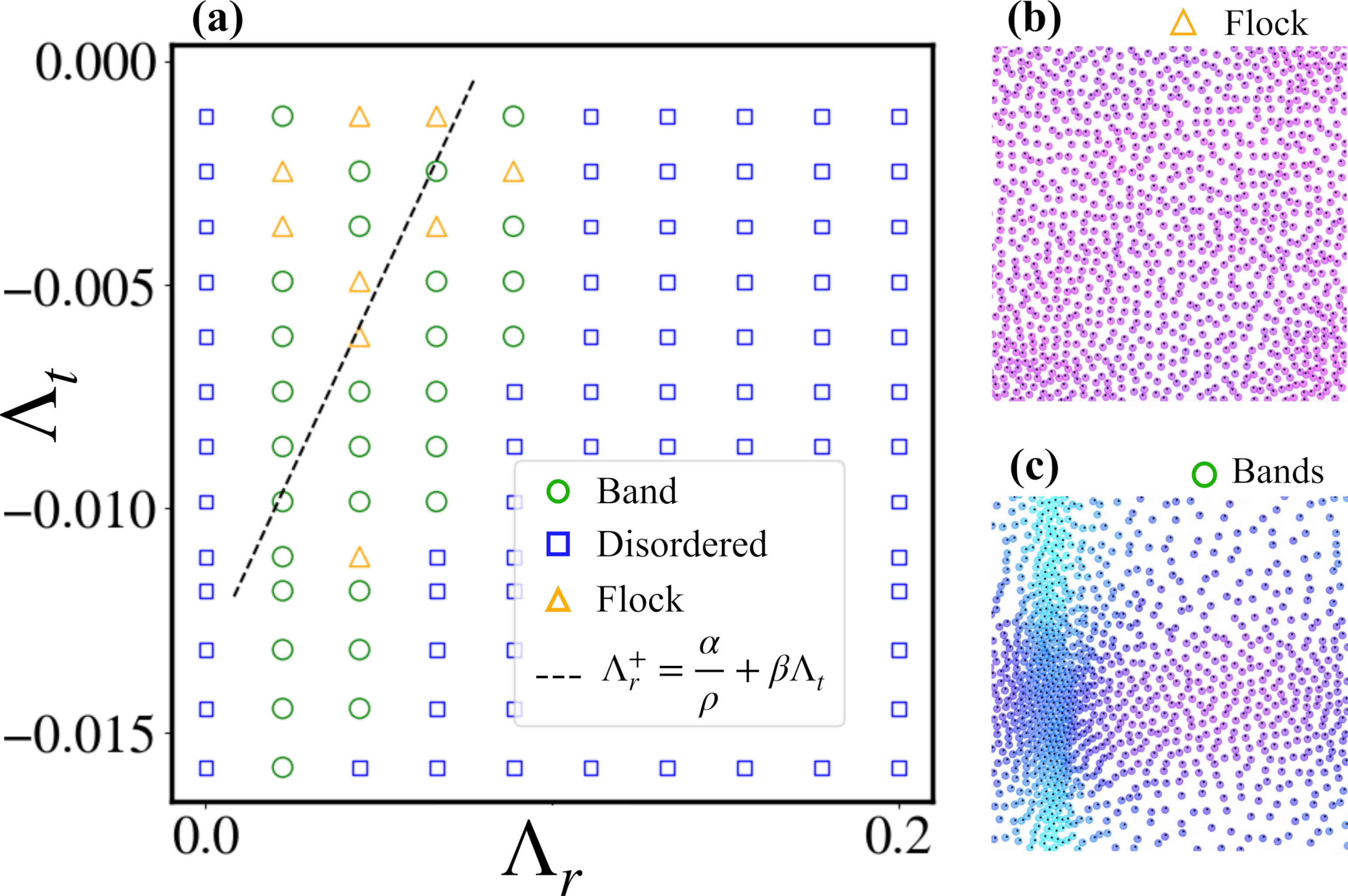}
    \caption{ 
     (a) Phase diagram in the $(\Lambda_r, \Lambda_t)$ plane for the case of $\chi_t < 0$. The transition line is extrapolated from (\ref{fig_CCF})(a). (b) liquid flock, (c) density band representing the phases of the phase diagram
      (same color scheme as in Fig (\ref{fig_CLF})). Total number of particles $N=1024$. 
    }
\label{fig:chi_t_attractive}
\end{figure}

\section{Effect of long-ranged translational chemo-attraction}
\label{sec:chi_t_pos}

\textcolor{black}{To further support the robustness of the mechanism, we report flocks arising from repulsive torques with the addition of long-ranged \textit{attractive} inter-particle forces, thus $\chi_t < 0$. We find that collective motion as flocks is maintained for $\Lambda_t \in [0,-0.01]$, before transitioning to a disordered state for larger attractive translational interactions; this is shown in Figure \ref{fig:chi_t_attractive}(a). In this regime, the clumping tendency is balanced by the excluded volume repulsions. This sets the y-intercept of the theoretical prediction in Fig. \ref{fig:chi_t_attractive}(a) to be of the same order of magnitude to the lower horizontal line in Fig. \ref{fig_CLF}. These are hence liquid flocks (CLF, see Fig. \ref{fig:chi_t_attractive}(b)). The band formation is also similar to that of the CLF. During density band formation, often less denser waves form in the system, and they move perpendicular to the direction of the density bands. Similar cross-sea phases are also observed in Vicsek-like models \cite{kursten2020dry}.}


\section{Continuum description}\label{sec:continuum}
To explain the above mechanism, we write top-down hydrodynamic equations for the density $\rho(\mathbf{r},t)$ and polarization $\mathbf{p}(\mathbf{r},t)$ fields by appealing to the symmetries and conservation laws of the system  \cite{chaikin1995principles}.
The continuum equations for our system can be written as:
\begin{subequations}
\label{eq:hydro}
\begin{align}
        \partial_t \rho =& -\bm\nabla\cdot \left( v_{\text{tr}} [\rho]\,\mathbf{p} - D_{\text{tr}}[\rho]\,\bm\nabla \rho \right) 
        \\   
        \partial_t \mathbf{p} =\,&A[\rho] \frac{\nabla \rho}{2}   + \frac{D_{\text{tr}}[\rho]}{\rho}\Big[\nabla \cdot (\mathbf{p} \nabla \rho) \Big].
    \end{align}
 \end{subequations}
Here, we have defined:
    \begin{align*}
        v_{\text{tr}}[\rho] &= v_{0} - \chi_t \xi_0 \rho, \quad
        D_{\text{tr}}[\rho] = \xi_t \rho \chi_t, \\
        A[\rho] &= (\xi_r \chi_r + \xi_0 \chi_t) \rho-   v_{\text{tr}}[\rho]. 
    \end{align*}
\textcolor{black}{The term $v_{\mathrm{tr}}$ (a coarse-grained speed field) appears in both the cross-coupling terms between density and polarization fields, stabilizing the density field at larger $\chi_t$. This effect is in turn countered by $D_{tr}$ (coarse-grained diffusivity), which randomizes the fields at large $\chi_t$. The $\chi_r$ and $\chi_t$ contributions via the $A$ term acts to destabilize the polarization field with respect to density gradients. Given that the transition lines observed are dependent on the area fraction simulated, each of these terms have an explicit density dependence via the constants $\xi_0$, $\xi_t$, and $\xi_r$, which are obtained by fitting the transition line predicted by the particle-based simulation in \ref{fig_CCF}(a). A route to these equations via a systematic coarse-grained procedure is well established \cite{grossmann2020particle, zhang2021active}, though it is not pursued in this work.  
We note that there are no stochastic contributions appearing in the dynamics of the fields $\rho$ and $\mathbf{p}$ in Eq.\eqref{eq:hydro}. We have assumed noise to be negligibly small as explained in the previous section. }
In effect, the presence of these stochastic terms in the particle-based model is merely necessitated for the formal coarse-graining procedure, but do not play any further role in explaining the instability of the flocking state. \\

\subsection{Case of $\chi_t = 0$}
First, we may consider the case of only rotational torques. See Fig. \ref{fig_CLF}(a) for phenomenology of this case with $\chi_t=0$ from the particle-based model.
From (\ref{eq:hydro}) - on setting $\chi_t=0$ - we obtain an enslaved dynamics of $\mathbf{\dot{p}}$ with respect to the density gradients:
\begin{align}
    \partial_t \mathbf{p} = \frac{\nabla \rho}{2} \Big[ (\chi_r \xi_r \rho - v_{0}) \Big].
\label{si_eq:enslaved}
\end{align}
From the above, we see that for the temporal evolution of $\mathbf{p}$ to change the nature of stability with respect to the density gradients (feedback response), we require the following:
\begin{align}
    \rho^{*}= \frac{v_{0}}{\chi_r\xi_r }.
\label{eq:rho_Star}
\end{align}
We see that the fluctuations in both $\rho$ and $\mathbf{p}$ are strongly coupled and that any fluctuations are stabilized in the long-time limit.

\subsection{General case with $\chi_t \neq 0$}
For the generic case of both $\chi_r\neq 0$ and $\chi_t\neq0$, we may note that solving for $\dot{\mathbf{p}} = 0$ in (\ref{eq:hydro}), whilst ignoring fluctuations, we obtain:
$$\rho^{*} = \frac{v_{0}}{\xi_r \chi_r + 2\xi_0 \chi_t}.
$$
thus generalizing \eqref{eq:rho_Star} for $\chi_t\neq 0$. In the following, we study the stability of the system about these steady-state values for the density field $\rho^*$ and a steady-state value of polar order $\mathbf{p}^*$. 

\subsection{Stability of the flocking state}
To study the stability of the flocking state, we linearize (\ref{eq:hydro}) about the steady-state density and polarizations ($\rho^{*},\mathbf{p}^{*}$), defining $(\delta \rho, \delta \mathbf{p}) = (\rho - \rho^{*}, \mathbf{p} - \mathbf{p}^{*})$. We use the following choice of Fourier basis expansion
\begin{subequations}
    \begin{align}
        \delta \tilde{\rho}(\mathbf{q},\omega) &= \int e^{i \mathbf{q} \cdot \mathbf{r}} e^{i \omega t} \delta \rho(\mathbf{r},t) d^{2} \mathbf{r} \\
        \delta \tilde{\theta}(\mathbf{q},\omega) &= \int e^{i \mathbf{q} \cdot \mathbf{r}} e^{i \omega t} (\nabla \cdot \delta \mathbf{p}) d^{2} \mathbf{r} 
    \end{align}
\label{si_eq:fourier_exp}
\end{subequations}
For a phase with uniform global polarization, the following conditions will hold: $\nabla \rho^{*} \rightarrow 0$ and $\nabla \cdot \mathbf{p}^{*} \rightarrow 0$.\\

Eq. (\ref{eq:hydro}) can then be linearized in the bases of (\ref{si_eq:fourier_exp}) to obtain the stability condition for the flocking state and subsequently derive the boundary line of Eq. (\ref{eq:ineq_chit_chir}), for a generic $\chi_t$.
The full linearization in Fourier space yields the following:

\begin{align}
      \frac{d}{dt}  \begin{pmatrix}
    \delta \tilde{\rho}\\\\
        \delta \tilde{\theta}
    \end{pmatrix}
    &= 
    \begin{pmatrix}
        -q^{2} \xi_{t} \chi_t \rho^{*} & -v_{\text{tr}}[\rho^{*}]\, 
        \\\\
        -{q^{2}} A[\rho^{*}] /2& 0
    \end{pmatrix}
    \begin{pmatrix}
        \delta \tilde{\rho} 
        \\\\
        \delta \tilde{\theta}
    \end{pmatrix}
\label{si_eq:linearized}
\end{align}
\\
Here $A[\rho^{*}] = (2\xi_0 \chi_t + \xi_r \chi_r)\rho^{*} - v_{0} $. 
The dispersion relation for $\omega(q)$ then reads
\begin{align}
    \omega = \frac{iq^{2} \xi_r\chi_t\rho^{*}}{2} \Big[ 1 \pm \sqrt{1 + \frac{2v_{\text{tr}}q^{2} A}{q^{4} \xi_t^{2} \chi_t^{2} {\rho^*}^{2} } } \Big]
\label{si_eq:disp}
\end{align}
with the condition for linear instability given by:
$$
    2\xi_0^{2} {\rho^*}^2 \chi_t^2 - \xi_0 \rho^{*} (2v_{0} - \xi_r \chi_r {\rho^*})\chi_t - v_{0}(\xi_r \chi_r \rho^* - v_{0}) < 0$$
Using the above, we can solve for $\chi_t$. For the value:
\begin{align}
    \chi_t^{\pm} = \frac{v_{0} - \frac{\xi_r \chi_r \rho^*}{2}}{2 \xi_0 {\rho^*}} \Big[ 1 \pm \sqrt{1 + \frac{2(\frac{\xi_r \chi_r \rho^{*}}{v_{0}} - 1)}{(1 - \frac{\xi_r \chi_r \rho^{*}}{2v_{0}})^{2}}} \Big]
\label{si_eq:quad_chitcrit}
\end{align}
we can consider the solution $\chi_t > \chi_t^{-}$. Let us further study the limit $\frac{\chi_r}{v_{0}} << 1$, which is corresponds to parameter ranges used in our simulations (see Table \ref{si_table_params}). \\

Using the definitions for $\Lambda_r $ and $\Lambda_t$ - see Eq.\eqref{eq:dimNumbs} - 
the condition for instability of the modes $\omega(q)$ is given by
\begin{align}
    \Lambda_r^{*} \gtrsim  \frac{2}{3 b^3 \xi_r\rho^{*}} - \frac{1.67 b \xi_0}{\xi_r} \Lambda_t
\label{si_eq:ineq_lr}
\end{align}
We can identify this to have the same functional form as Eq.(\ref{eq:ineq_chit_chir}) defining the transition line.
Equating with Eq. (\ref{eq:ineq_chit_chir}), we have that
    \begin{align}
        \alpha &= \frac{2}{3 b^3\xi_r},\quad \quad
        \beta = -\frac{1.67b \xi_0}{\xi_r} 
\label{sI_eq:const_eq}
    \end{align}
We can thus infer the values of $\xi_r \approx 2/(3 b^3\alpha) \approx 51.3$ and $\xi_0 \approx -5.13/b$. The hydrodynamic instability of the flocking phase thus captures the transition lines between ordered and disordered phase in Figures \ref{fig_CLF}(a) and \ref{fig_CCF}(a). Though Eq. (\ref{si_eq:ineq_lr}) captures the transition for a generic crystalline flock (generic $\chi_t$), we note that there is no explicit spatial-structure instability encoded in the hydrodynamic model \cite{geyer2019freezing}, thus if one simulates Eq. (\ref{eq:hydro}) one obtains effectively fluidic flocks at a continuum level. Nevertheless, we conclude that the \textit{noiseless} limit of a minimal hydrodynamic model captures the destabilization transition of the flock, with exact details of the spatial structure of the system likely requiring to go beyond the conventional molecular chaos hypothesis \cite{chen2023molecular, grossmann2020particle}. \\

\section{Summary and discussion}\label{sec:conc}
To conclude, we present a minimal mechanism for formation of fluid flocks that only requires long-ranged (net) repulsive torques and short-ranged repulsive forces from
steric interactions. 
Although the roles of excluded volume interactions \cite{martin2018collective, caprini2023flocking, knevzevic2022collective, chen2023molecular}, repulsive torques - either short \cite{knevzevic2022collective} or long \cite{ das2024flocking} ranged - and long-ranged repulsive forces \cite{caprini2023flocking, das2024flocking} have been studied separately, we have specified here the individual contributions of each of those components, and argued that the former two are the minimal requisite components. The mechanism can summarized as follows: the pair colloids slide together for at least a unit length before deterministically rotating away from each other. This is required to break the forward-backward symmetry at the pair-collision level. Destabilizing of the flock arises due to symmetric collisions at the pair colloidal level, wherein they slide exceedingly short before rotating away upon collision. These results we show to be consistent with the noiseless limit of a minimal hydrodynamic theory. One may ask whether such a sliding mechanism alone added to cognitive based (agent based) models \cite{adhikary2022pattern, adhikary2023collective} could reproduce the flocking transition; indeed we will leave this to future work.\\

\textcolor{black}{It is important to note that the CLF and CCF phases are both generated by, and further destabilized by, the deterministic torques
themselves, independent of the contribution of noise. Though we have explicit noise terms in the updates in Eq.(\ref{eq:mainLE}) for the sake of completion, they are set to be negligibly small (see Table \ref{si_table_params} in Appendix).  
We note that we have repeated all the simulations presented in Fig. (\ref{fig_CLF}) with $D_r=D_t=0$, and have found the results to be identical. 
Thus, our results on flocking of phoretically interacting active particles and their destabilization are distinct from  conventional picture of noise-induced inhibition of flock formation \cite{chate2008collective, adhikary2022pattern, caprini2023flocking, das2024flocking}.}
We note that in the absence of short-ranged excluded volume repulsions, 
our results on the formation of CLF would converge to that recently studied in \cite{das2024flocking} 
where the need for (long-ranged) repulsive forces is more imminent. Translational repulsion merely renders the flock to acquire a crystalline structure.  We also find flocks for (a narrow range of) long-ranged \textit{attractive} forces, if of the order of less than the short-ranged repulsion. 
 We have not studied here the role of particle shape anisotropy \cite{grossmann2020particle, wensink2012emergent}, which would add additional competing length scales to our analysis. We have also not studied the role of hydrodynamic interactions between the particles \cite{marchetti2013hydrodynamics, aditi2002hydrodynamic, thutupalli2018flow}. We expect our results here to be directly relevant to a variety of experimental systems. For example, migrating cell collections have been widely reported to exhibit spatial organization resembling a polar fluid \cite{hayakawa2020polar, hiraiwa2020dynamic}, while their interaction mechanism has been known to include (among others) long-ranged chemotaxis and avoidance torques \cite{camley2017physical}. The presented mechanism being very generic, we expect it to also be relevant for interacting colloids with non-diffusiophoretic fields \cite{bricard2013emergence, zhang2021active, geyer2019freezing}, where the underlying mechanism should also hold. 

\section*{Conflicts of interest}
There are no conflicts to declare.
\section*{Data availability}
The datasets are generated from simulation. They are available from the corresponding authors on reasonable request.
 
\section*{Acknowledgments}
We thank Professor Mike Cates for discussions.
\textcolor{black}{We also extend our gratitude to two anonymous reviewers of the manuscript for their feedback and constructive criticism, which led to
an improvement in the presentation of our results.} AGS acknowledges funding from the DIA Fellowship from the Government of India. SA acknowledges support from the National Postdoctoral Fellowship (SERB File number: PDF/2023/002096) provided by ANRF, Government of India. RS acknowledges support from seed and initiation grants from IIT Madras as well as a Start-up Research Grant from SERB (SERB File Number: SRG/2022/000682), India.

\appendix
\section*{Appendices}

\subsection{Table of parameters}
A table of all the parameters used for the generated figures is presented in Table \ref{si_table_params}. \\

\begin{table*}[t!]
	\centering
	\begin{tabular}[c]{|l|l|l|l|l|l|l|l|l|l|l|l|l|l|l}
     	\hline
     	Figure no. & $L$ &$dt$ & $b$ & $v_{0}$  & $\kappa$ & $N$ & $\phi$ & $\chi_t$ & $\Lambda_t$ & $\chi_r$ & $\Lambda_r$ & $D_r$ & $D_t$ \\
     	\hline
     	2 & $100$ & $0.01$ & $1$ &  $50 $ & $175 $ & $(319,2578)$ &  $(0.1,0.8)$ & $0$ & $0$ & $(0,7)$ & $(0.0,0.14)$ & $ 10^{-6}$ & $10^{-8}$ \\
        \hline
     	3.(a) & $100$ & $0.01$ & $1$ & $50 $ & $175 $ & $1024$ & $0.32$ & $(0,5)$ & $(0.0,0.1)$ & $(0,10)$ & $(0.0,0.2)$ & $ 10^{-6}$ & $10^{-8}$ \\
        \hline
        4.(a),(c) & $186$ &$0.01$ & $1$ &   $50 $ & $175 $ & $4000$ & $0.36$ & $0$ & $0$ & $0.75$ & $0.015$ & $ 10^{-6}$ & $10^{-8}$ \\
        \hline
        4.(b),(c) & $186$ &$0.01$ & $1$ &   $50 $ & $175 $ & $4000$ & $0.36$ & $3$ & $0.06$ & $6$ & $0.12$ & $10^{-6}$ & $10^{-8}$ \\
        \hline     
        5.(a) & $(77,354)$ &$0.01$ & $1$ &  $50 $ & $175 $ & $1600$ & $(0.04,0.84)$ & $0$ & $0$ & $(0,9)$ & $(0,0.18)$ & $ 10^{-6}$ & $10^{-8}$ \\
        \hline
        5.(b) & $118$ & $0.01$ & $1$ &  $50$ & $175 $ & $1600$ & $0.36$ & $(0,12)$ & $(0,0.6)$ & $(0,32)$ & $(0,0.64)$ & $10^{-6}$ & $10^{-8}$ \\
        \hline    
        6.(a) & $100$ &$0.01$ & $1$ & $50 $ & $175 $ & $(637,1592)$ & $(0.2,0.5)$ & $0$ & $0$ &$(0,8)$ & $(0.0,0.16)$  & $ 10^{-6}$ & $10^{-8}$ \\
        \hline
        6.(c) & $100$ &$0.01$ & $1$ &   $50 $ & $175 $ & $(637,1592)$ & $(0.2,0.5)$ & $(0,10)$ & $(0.0,0.2)$ & $27$ & $0.54$ & $10^{-6}$ & $10^{-8}$ \\
        \hline 
        7.(a) & $100$ &$0.01$ & $1$ & $50 $ & $175 $ & $1024$ & $0.32$ & $(-1,0)$ & $(-0.02,0.0)$ & $(0,10)$ & $(0.0,0.2)$ & $10^{-6}$ & $10^{-8}$ \\
        \hline
	\end{tabular}
 \caption{Parameter values used for respective figures of the paper. 
 Here $L$ is system size, $b$ is the radius of the particle, $N$ is the number of particles, $dt$ is the time-stepping of the Euler-Maruyama integrator. The remaining parameters are defined after Eq.\eqref{eq:mainLE}. We have kept $\frac{c_0}{4\pi D_c}=1$ in all the simulations. Note that we consider noise to be less dominant to deterministic effects. Indeed, for the results in this paper, the Péclet number $\mathrm{Pe} = v_{0}/(bD_r)$ is around $10^9$. }
 \label{si_table_params}
\end{table*}


%
\section{Pair correlation functions}\label{app:CCFCLF} 
The pair distribution function $g(r,\varphi,\theta)$ implies correlations and provides the probability density of finding a pair of particles at a distance $r$ with self-propulsion angle $\theta$. 
Here, $\varphi$ is the angle between the relative position $r$ and the self-propulsion direction of the tagged particle, $\hat{\mathbf{e}}_i \cdot \hat{\mathbf{{r}}}_{ij} = \cos\varphi$ and $\theta$ is the angle between the propulsion directions of the particle pair. To obtain this pair distribution function, histograms are created with the number of particles $N_h(r,\varphi,\theta)$ at distance $r$, positional angle $\varphi$ and angle $\theta$. The bins are chosen as $0.1 b$ for $r$, $\pi/150$ for $\varphi$ and $\pi/2$ for $\theta$. Then we need to normalize $N_h(r,\varphi,\theta)$, to obtain the pair distribution function $g(r,\varphi,\theta)$. The number of configurations is $t_c$($10000$ snapshots), over which histograms are made. The area of the annular segment of radial width $dr$, angular width $d\varphi$ is given by $A_r=r dr d\varphi$. The number density is $\rho=N/L^{2}$ and the size of the relative orientation bin is $d\theta$. Then the normalization factor is $N_r=\frac{A_{r}\rho N t_c d\theta}{2\pi}$ and the pair distribution function is given by $g(r,\varphi,\theta)=N_h(r,\varphi,\theta)/N_r$.\\
 
\begin{figure}
    \centering
    \includegraphics[width=0.45\textwidth]{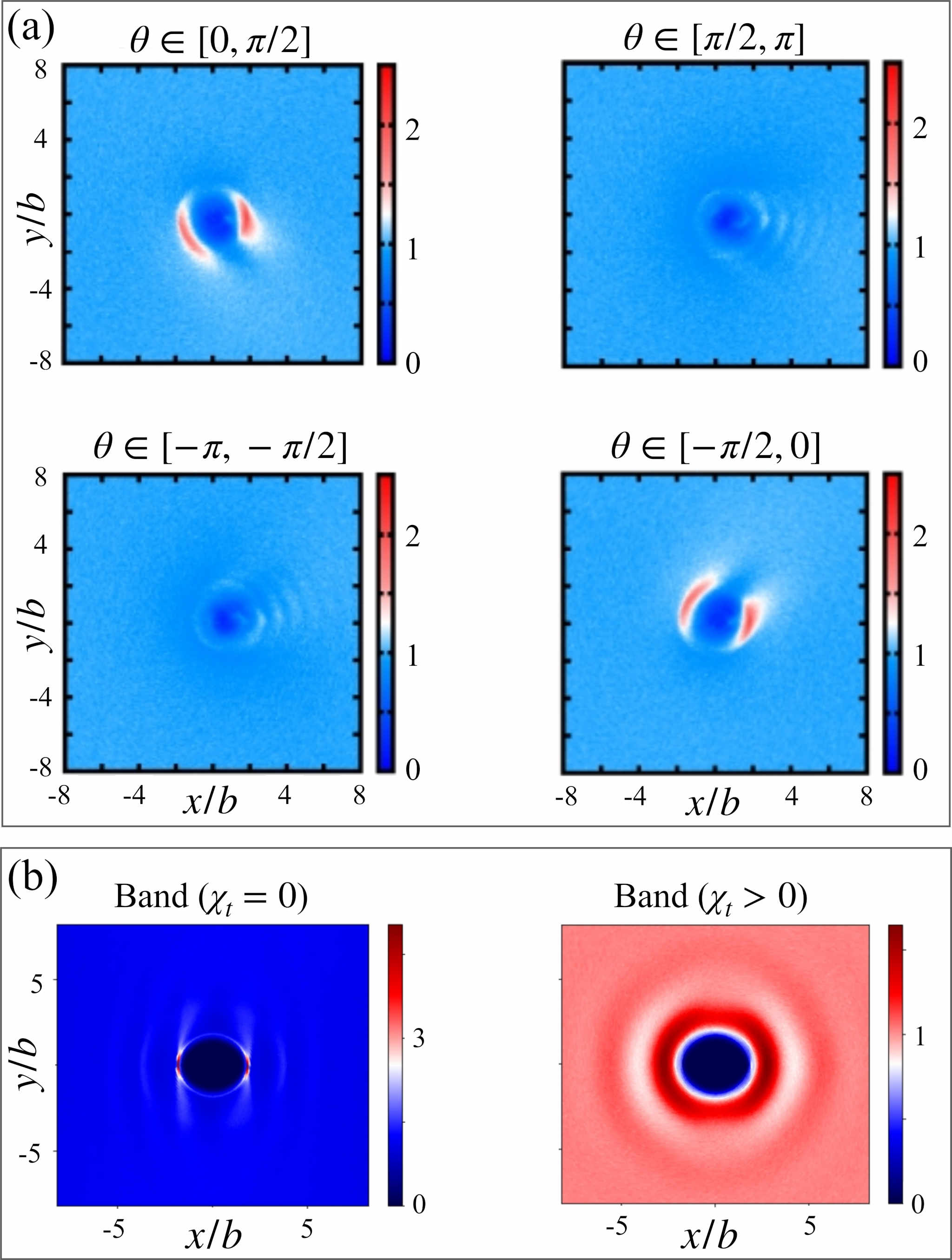}
    \caption{Pair correlation function $g(r,\varphi,\theta)$ for the isotropic phase for four different $\theta$ ranges in panel (a). In panel (b), we show pair correlation function $g(r,\varphi)$ in the band phase for $\chi_t=0$  and $\chi_t>0$. }
\label{fig:grIB}
\end{figure}

An example of this for isotropic phase is given in Fig. \ref{fig:grIB}(a). The pair distribution functions are determined for the four different $\theta$ ranges as: $\theta$:[$0,\pi/2$], $\theta$:[$\pi/2,\pi$], $\theta$:[$-\pi, -\pi/2$] and $\theta$:[$-\pi/2,0$]. We see that in the isotropic phase, the probability of finding a neighboring particle is different in the front relative to the back of the direction of motion (orientation), indicative of the symmetry breaking in an active disordered gas \cite{grossmann2020particle, zhang2021active, das2024flocking}. In the main text, we show that $g(r,\varphi)$ also distinguishes CLFs from CCFs, the former having a four-lobe structure compared to the latter's hexatic structure. In addition, we compare the structure of $g(r,\varphi)$ for the band phases in Fig. (\ref{fig:grIB})(b), for the CLF and CCF density bands. We see that the CLF and CCF bands are distinguished strongly via this measure: in the former there is a strong likelihood of finding neighbouring colloids in either the front or back of the propulsion direction, whilst in the latter the aggregate structure resembles that of a (non-polar) liquid due to the presence of cross bands. Note that the band structure is more dense in the case of $\chi_t=0$, resulting a more intense correlation in $g(r,\varphi)$. A related quantity that we compute is the radial $g(r)$.
This is plotted in Fig. (\ref{fig:CCF-CLF_gr}), distinguishing the CLF and CCF.

\section{Description of the supplementary movies}
The time evolution of the CCF and CLF structures for a larger system with total number of
particles $N=20000$, are attached as supplementary movies \cite{siText}. We note that our results 
are robust as we change number of particles in the simulations. 
For both cases, we
started with random initial conditions (positions and orientations), and
update the equation of motion following Eq. \ref{eq:mainLE}.
In both movies, the particles are colored  by their orientations (given in terms of the angle $\theta$ their orientation vector makes with the positive $x$-axis). The color bar is same as the one used in Fig. \ref{fig_CLF}.
Two particles are intentionally colored black throughout the simulation to simply trace them during
the dynamics. Polar order or the flocking state collectively emerges in this model with both the cases. Parameters used for the two movies are below.
\begin{itemize}
    \item \textbf{Movie I}.
In the case of $\chi_t = 0$ (CLF), the liquid flock gradually forms throughout the system and moves
in a particular direction. The fixed parameter values are: $L=418$, $b=1$, $dt=0.01$, $D_r=10^{-3}$,
$D_t=10^{-3}$, $\kappa=175$, $v_{0}=50$, $\chi_t=0$, $\chi_r=0.75$.
\item \textbf{Movie II}.
In the case of $\chi_t > 0$ (CCF), the particles first start to form local polar flocks (colored patch),
then merge and pick a particular direction of motion entirely. The fixed parameter values are:
$L=418$, $b=1$, $dt=0.01$, $D_r=10^{-3}$, $D_t=10^{-3}$, $\kappa=175$, $v_{0}=50$, $\chi_t=3$, $\chi_r=6$.
\item \textbf{Movie III}.
\textcolor{black}{In the case of $\chi_t = 0$ (CLF), particles form several density bands through out the space. Additionally, less dense waves (which move perpendicular to the direction of density bands) are also seen to form with these density bands. The fixed parameter values are:
$L=418$, $b=1$, $dt=0.01$, $D_r=10^{-3}$, $D_t=10^{-3}$, $\kappa=175$, $v_{0}=50$, $\chi_t=0$, $\chi_r=2.0$.}
\end{itemize}

\begin{thebibliography}{51}%
\makeatletter
\providecommand \@ifxundefined [1]{%
 \@ifx{#1\undefined}
}%
\providecommand \@ifnum [1]{%
 \ifnum #1\expandafter \@firstoftwo
 \else \expandafter \@secondoftwo
 \fi
}%
\providecommand \@ifx [1]{%
 \ifx #1\expandafter \@firstoftwo
 \else \expandafter \@secondoftwo
 \fi
}%
\providecommand \natexlab [1]{#1}%
\providecommand \enquote  [1]{``#1''}%
\providecommand \bibnamefont  [1]{#1}%
\providecommand \bibfnamefont [1]{#1}%
\providecommand \citenamefont [1]{#1}%
\providecommand \href@noop [0]{\@secondoftwo}%
\providecommand \href [0]{\begingroup \@sanitize@url \@href}%
\providecommand \@href[1]{\@@startlink{#1}\@@href}%
\providecommand \@@href[1]{\endgroup#1\@@endlink}%
\providecommand \@sanitize@url [0]{\catcode `\\12\catcode `\$12\catcode `\&12\catcode `\#12\catcode `\^12\catcode `\_12\catcode `\%12\relax}%
\providecommand \@@startlink[1]{}%
\providecommand \@@endlink[0]{}%
\providecommand \url  [0]{\begingroup\@sanitize@url \@url }%
\providecommand \@url [1]{\endgroup\@href {#1}{\urlprefix }}%
\providecommand \urlprefix  [0]{URL }%
\providecommand \Eprint [0]{\href }%
\providecommand \doibase [0]{https://doi.org/}%
\providecommand \selectlanguage [0]{\@gobble}%
\providecommand \bibinfo  [0]{\@secondoftwo}%
\providecommand \bibfield  [0]{\@secondoftwo}%
\providecommand \translation [1]{[#1]}%
\providecommand \BibitemOpen [0]{}%
\providecommand \bibitemStop [0]{}%
\providecommand \bibitemNoStop [0]{.\EOS\space}%
\providecommand \EOS [0]{\spacefactor3000\relax}%
\providecommand \BibitemShut  [1]{\csname bibitem#1\endcsname}%
\let\auto@bib@innerbib\@empty
\bibitem [{\citenamefont {Toner}(2024{\natexlab{a}})}]{toner2024physics}%
  \BibitemOpen
  \bibfield  {author} {\bibinfo {author} {\bibfnamefont {J.}~\bibnamefont {Toner}},\ }\href@noop {} {\emph {\bibinfo {title} {The Physics of Flocking: Birth, Death, and Flight in Active Matter}}}\ (\bibinfo  {publisher} {Cambridge University Press},\ \bibinfo {year} {2024})\BibitemShut {NoStop}%
\bibitem [{\citenamefont {Couzin}\ \emph {et~al.}(2002)\citenamefont {Couzin}, \citenamefont {Krause}, \citenamefont {James}, \citenamefont {Ruxton},\ and\ \citenamefont {Franks}}]{couzin2002collective}%
  \BibitemOpen
  \bibfield  {author} {\bibinfo {author} {\bibfnamefont {I.~D.}\ \bibnamefont {Couzin}}, \bibinfo {author} {\bibfnamefont {J.}~\bibnamefont {Krause}}, \bibinfo {author} {\bibfnamefont {R.}~\bibnamefont {James}}, \bibinfo {author} {\bibfnamefont {G.~D.}\ \bibnamefont {Ruxton}},\ and\ \bibinfo {author} {\bibfnamefont {N.~R.}\ \bibnamefont {Franks}},\ }\bibfield  {title} {\bibinfo {title} {Collective memory and spatial sorting in animal groups},\ }\href@noop {} {\bibfield  {journal} {\bibinfo  {journal} {Journal of theoretical biology}\ }\textbf {\bibinfo {volume} {218}},\ \bibinfo {pages} {1} (\bibinfo {year} {2002})}\BibitemShut {NoStop}%
\bibitem [{\citenamefont {Marchetti}\ \emph {et~al.}(2013)\citenamefont {Marchetti}, \citenamefont {Joanny}, \citenamefont {Ramaswamy}, \citenamefont {Liverpool}, \citenamefont {Prost}, \citenamefont {Rao},\ and\ \citenamefont {Simha}}]{marchetti2013hydrodynamics}%
  \BibitemOpen
  \bibfield  {author} {\bibinfo {author} {\bibfnamefont {M.~C.}\ \bibnamefont {Marchetti}}, \bibinfo {author} {\bibfnamefont {J.-F.}\ \bibnamefont {Joanny}}, \bibinfo {author} {\bibfnamefont {S.}~\bibnamefont {Ramaswamy}}, \bibinfo {author} {\bibfnamefont {T.~B.}\ \bibnamefont {Liverpool}}, \bibinfo {author} {\bibfnamefont {J.}~\bibnamefont {Prost}}, \bibinfo {author} {\bibfnamefont {M.}~\bibnamefont {Rao}},\ and\ \bibinfo {author} {\bibfnamefont {R.~A.}\ \bibnamefont {Simha}},\ }\bibfield  {title} {\bibinfo {title} {Hydrodynamics of soft active matter},\ }\href@noop {} {\bibfield  {journal} {\bibinfo  {journal} {Reviews of modern physics}\ }\textbf {\bibinfo {volume} {85}},\ \bibinfo {pages} {1143} (\bibinfo {year} {2013})}\BibitemShut {NoStop}%
\bibitem [{\citenamefont {Vicsek}\ and\ \citenamefont {Zafeiris}(2012)}]{vicsek2012collective}%
  \BibitemOpen
  \bibfield  {author} {\bibinfo {author} {\bibfnamefont {T.}~\bibnamefont {Vicsek}}\ and\ \bibinfo {author} {\bibfnamefont {A.}~\bibnamefont {Zafeiris}},\ }\bibfield  {title} {\bibinfo {title} {Collective motion},\ }\href@noop {} {\bibfield  {journal} {\bibinfo  {journal} {Physics reports}\ }\textbf {\bibinfo {volume} {517}},\ \bibinfo {pages} {71} (\bibinfo {year} {2012})}\BibitemShut {NoStop}%
\bibitem [{\citenamefont {Vicsek}\ \emph {et~al.}(1995)\citenamefont {Vicsek}, \citenamefont {Czir{\'o}k}, \citenamefont {Ben-Jacob}, \citenamefont {Cohen},\ and\ \citenamefont {Shochet}}]{vicsek1995novel}%
  \BibitemOpen
  \bibfield  {author} {\bibinfo {author} {\bibfnamefont {T.}~\bibnamefont {Vicsek}}, \bibinfo {author} {\bibfnamefont {A.}~\bibnamefont {Czir{\'o}k}}, \bibinfo {author} {\bibfnamefont {E.}~\bibnamefont {Ben-Jacob}}, \bibinfo {author} {\bibfnamefont {I.}~\bibnamefont {Cohen}},\ and\ \bibinfo {author} {\bibfnamefont {O.}~\bibnamefont {Shochet}},\ }\bibfield  {title} {\bibinfo {title} {Novel type of phase transition in a system of self-driven particles},\ }\href@noop {} {\bibfield  {journal} {\bibinfo  {journal} {Physical review letters}\ }\textbf {\bibinfo {volume} {75}},\ \bibinfo {pages} {1226} (\bibinfo {year} {1995})}\BibitemShut {NoStop}%
\bibitem [{\citenamefont {Chat{\'e}}\ \emph {et~al.}(2008)\citenamefont {Chat{\'e}}, \citenamefont {Ginelli}, \citenamefont {Gr{\'e}goire},\ and\ \citenamefont {Raynaud}}]{chate2008collective}%
  \BibitemOpen
  \bibfield  {author} {\bibinfo {author} {\bibfnamefont {H.}~\bibnamefont {Chat{\'e}}}, \bibinfo {author} {\bibfnamefont {F.}~\bibnamefont {Ginelli}}, \bibinfo {author} {\bibfnamefont {G.}~\bibnamefont {Gr{\'e}goire}},\ and\ \bibinfo {author} {\bibfnamefont {F.}~\bibnamefont {Raynaud}},\ }\bibfield  {title} {\bibinfo {title} {Collective motion of self-propelled particles interacting without cohesion},\ }\href@noop {} {\bibfield  {journal} {\bibinfo  {journal} {Physical Review E—Statistical, Nonlinear, and Soft Matter Physics}\ }\textbf {\bibinfo {volume} {77}},\ \bibinfo {pages} {046113} (\bibinfo {year} {2008})}\BibitemShut {NoStop}%
\bibitem [{\citenamefont {Mishra}\ \emph {et~al.}(2010)\citenamefont {Mishra}, \citenamefont {Baskaran},\ and\ \citenamefont {Marchetti}}]{mishra2010fluctuations}%
  \BibitemOpen
  \bibfield  {author} {\bibinfo {author} {\bibfnamefont {S.}~\bibnamefont {Mishra}}, \bibinfo {author} {\bibfnamefont {A.}~\bibnamefont {Baskaran}},\ and\ \bibinfo {author} {\bibfnamefont {M.~C.}\ \bibnamefont {Marchetti}},\ }\bibfield  {title} {\bibinfo {title} {Fluctuations and pattern formation in self-propelled particles},\ }\href@noop {} {\bibfield  {journal} {\bibinfo  {journal} {Physical Review E—Statistical, Nonlinear, and Soft Matter Physics}\ }\textbf {\bibinfo {volume} {81}},\ \bibinfo {pages} {061916} (\bibinfo {year} {2010})}\BibitemShut {NoStop}%
\bibitem [{\citenamefont {Toner}\ \emph {et~al.}(2005)\citenamefont {Toner}, \citenamefont {Tu},\ and\ \citenamefont {Ramaswamy}}]{toner2005hydrodynamics}%
  \BibitemOpen
  \bibfield  {author} {\bibinfo {author} {\bibfnamefont {J.}~\bibnamefont {Toner}}, \bibinfo {author} {\bibfnamefont {Y.}~\bibnamefont {Tu}},\ and\ \bibinfo {author} {\bibfnamefont {S.}~\bibnamefont {Ramaswamy}},\ }\bibfield  {title} {\bibinfo {title} {Hydrodynamics and phases of flocks},\ }\href@noop {} {\bibfield  {journal} {\bibinfo  {journal} {Annals of Physics}\ }\textbf {\bibinfo {volume} {318}},\ \bibinfo {pages} {170} (\bibinfo {year} {2005})}\BibitemShut {NoStop}%
\bibitem [{\citenamefont {Toner}(2024{\natexlab{b}})}]{toner2024_BD}%
  \BibitemOpen
  \bibfield  {author} {\bibinfo {author} {\bibfnamefont {J.}~\bibnamefont {Toner}},\ }\bibfield  {title} {\bibinfo {title} {Birth, death, and horizontal flight: Malthusian flocks with an easy plane in three dimensions},\ }\href {https://doi.org/10.1103/PhysRevE.110.064604} {\bibfield  {journal} {\bibinfo  {journal} {Phys. Rev. E}\ }\textbf {\bibinfo {volume} {110}},\ \bibinfo {pages} {064604} (\bibinfo {year} {2024}{\natexlab{b}})}\BibitemShut {NoStop}%
\bibitem [{\citenamefont {Ikeda}(2024)}]{ikeda2024minimum}%
  \BibitemOpen
  \bibfield  {author} {\bibinfo {author} {\bibfnamefont {H.}~\bibnamefont {Ikeda}},\ }\bibfield  {title} {\bibinfo {title} {Minimum scaling model and exact exponents for the nambu-goldstone modes in the vicsek model},\ }\href@noop {} {\bibfield  {journal} {\bibinfo  {journal} {Physical Review Letters}\ }\textbf {\bibinfo {volume} {133}},\ \bibinfo {pages} {258301} (\bibinfo {year} {2024})}\BibitemShut {NoStop}%
\bibitem [{\citenamefont {Chat{\'e}}\ and\ \citenamefont {Solon}(2024)}]{chate2024dynamic}%
  \BibitemOpen
  \bibfield  {author} {\bibinfo {author} {\bibfnamefont {H.}~\bibnamefont {Chat{\'e}}}\ and\ \bibinfo {author} {\bibfnamefont {A.}~\bibnamefont {Solon}},\ }\bibfield  {title} {\bibinfo {title} {Dynamic scaling of two-dimensional polar flocks},\ }\href@noop {} {\bibfield  {journal} {\bibinfo  {journal} {Physical Review Letters}\ }\textbf {\bibinfo {volume} {132}},\ \bibinfo {pages} {268302} (\bibinfo {year} {2024})}\BibitemShut {NoStop}%
\bibitem [{\citenamefont {Jentsch}\ and\ \citenamefont {Lee}(2024)}]{jentsch2024new}%
  \BibitemOpen
  \bibfield  {author} {\bibinfo {author} {\bibfnamefont {P.}~\bibnamefont {Jentsch}}\ and\ \bibinfo {author} {\bibfnamefont {C.~F.}\ \bibnamefont {Lee}},\ }\bibfield  {title} {\bibinfo {title} {New universality class describes vicsek’s flocking phase in physical dimensions},\ }\href@noop {} {\bibfield  {journal} {\bibinfo  {journal} {Physical Review Letters}\ }\textbf {\bibinfo {volume} {133}},\ \bibinfo {pages} {128301} (\bibinfo {year} {2024})}\BibitemShut {NoStop}%
\bibitem [{\citenamefont {Adhikary}\ and\ \citenamefont {Santra}(2022)}]{adhikary2022pattern}%
  \BibitemOpen
  \bibfield  {author} {\bibinfo {author} {\bibfnamefont {S.}~\bibnamefont {Adhikary}}\ and\ \bibinfo {author} {\bibfnamefont {S.}~\bibnamefont {Santra}},\ }\bibfield  {title} {\bibinfo {title} {Pattern formation and phase transition in the collective dynamics of a binary mixture of polar self-propelled particles},\ }\href@noop {} {\bibfield  {journal} {\bibinfo  {journal} {Physical Review E}\ }\textbf {\bibinfo {volume} {105}},\ \bibinfo {pages} {064612} (\bibinfo {year} {2022})}\BibitemShut {NoStop}%
\bibitem [{\citenamefont {Adhikary}\ and\ \citenamefont {Santra}(2023)}]{adhikary2023collective}%
  \BibitemOpen
  \bibfield  {author} {\bibinfo {author} {\bibfnamefont {S.}~\bibnamefont {Adhikary}}\ and\ \bibinfo {author} {\bibfnamefont {S.}~\bibnamefont {Santra}},\ }\bibfield  {title} {\bibinfo {title} {Collective dynamics and phase transition of active matter in presence of orientation adapters},\ }\href@noop {} {\bibfield  {journal} {\bibinfo  {journal} {arXiv preprint arXiv:2302.13035}\ } (\bibinfo {year} {2023})}\BibitemShut {NoStop}%
\bibitem [{\citenamefont {Baconnier}\ \emph {et~al.}(2025)\citenamefont {Baconnier}, \citenamefont {Dauchot}, \citenamefont {D{\'e}mery}, \citenamefont {D{\"u}ring}, \citenamefont {Henkes}, \citenamefont {Huepe},\ and\ \citenamefont {Shee}}]{baconnier2025self}%
  \BibitemOpen
  \bibfield  {author} {\bibinfo {author} {\bibfnamefont {P.}~\bibnamefont {Baconnier}}, \bibinfo {author} {\bibfnamefont {O.}~\bibnamefont {Dauchot}}, \bibinfo {author} {\bibfnamefont {V.}~\bibnamefont {D{\'e}mery}}, \bibinfo {author} {\bibfnamefont {G.}~\bibnamefont {D{\"u}ring}}, \bibinfo {author} {\bibfnamefont {S.}~\bibnamefont {Henkes}}, \bibinfo {author} {\bibfnamefont {C.}~\bibnamefont {Huepe}},\ and\ \bibinfo {author} {\bibfnamefont {A.}~\bibnamefont {Shee}},\ }\bibfield  {title} {\bibinfo {title} {Self-aligning polar active matter},\ }\href@noop {} {\bibfield  {journal} {\bibinfo  {journal} {Reviews of Modern Physics}\ }\textbf {\bibinfo {volume} {97}},\ \bibinfo {pages} {015007} (\bibinfo {year} {2025})}\BibitemShut {NoStop}%
\bibitem [{\citenamefont {Chen}\ \emph {et~al.}(2023)\citenamefont {Chen}, \citenamefont {Welch}, \citenamefont {Leishangthem}, \citenamefont {Ghosh}, \citenamefont {Zhang}, \citenamefont {Sun}, \citenamefont {Klukas}, \citenamefont {Tu}, \citenamefont {Cheng},\ and\ \citenamefont {Xu}}]{chen2023molecular}%
  \BibitemOpen
  \bibfield  {author} {\bibinfo {author} {\bibfnamefont {L.}~\bibnamefont {Chen}}, \bibinfo {author} {\bibfnamefont {K.~J.}\ \bibnamefont {Welch}}, \bibinfo {author} {\bibfnamefont {P.}~\bibnamefont {Leishangthem}}, \bibinfo {author} {\bibfnamefont {D.}~\bibnamefont {Ghosh}}, \bibinfo {author} {\bibfnamefont {B.}~\bibnamefont {Zhang}}, \bibinfo {author} {\bibfnamefont {T.-P.}\ \bibnamefont {Sun}}, \bibinfo {author} {\bibfnamefont {J.}~\bibnamefont {Klukas}}, \bibinfo {author} {\bibfnamefont {Z.}~\bibnamefont {Tu}}, \bibinfo {author} {\bibfnamefont {X.}~\bibnamefont {Cheng}},\ and\ \bibinfo {author} {\bibfnamefont {X.}~\bibnamefont {Xu}},\ }\bibfield  {title} {\bibinfo {title} {Molecular chaos in dense active systems},\ }\href@noop {} {\bibfield  {journal} {\bibinfo  {journal} {arXiv preprint arXiv:2302.10525}\ } (\bibinfo {year} {2023})}\BibitemShut {NoStop}%
\bibitem [{\citenamefont {Gro{\ss}mann}\ \emph {et~al.}(2020)\citenamefont {Gro{\ss}mann}, \citenamefont {Aranson},\ and\ \citenamefont {Peruani}}]{grossmann2020particle}%
  \BibitemOpen
  \bibfield  {author} {\bibinfo {author} {\bibfnamefont {R.}~\bibnamefont {Gro{\ss}mann}}, \bibinfo {author} {\bibfnamefont {I.~S.}\ \bibnamefont {Aranson}},\ and\ \bibinfo {author} {\bibfnamefont {F.}~\bibnamefont {Peruani}},\ }\bibfield  {title} {\bibinfo {title} {A particle-field approach bridges phase separation and collective motion in active matter},\ }\href@noop {} {\bibfield  {journal} {\bibinfo  {journal} {Nature communications}\ }\textbf {\bibinfo {volume} {11}},\ \bibinfo {pages} {5365} (\bibinfo {year} {2020})}\BibitemShut {NoStop}%
\bibitem [{\citenamefont {Hiraiwa}(2020)}]{hiraiwa2020dynamic}%
  \BibitemOpen
  \bibfield  {author} {\bibinfo {author} {\bibfnamefont {T.}~\bibnamefont {Hiraiwa}},\ }\bibfield  {title} {\bibinfo {title} {Dynamic self-organization of idealized migrating cells by contact communication},\ }\href@noop {} {\bibfield  {journal} {\bibinfo  {journal} {Physical Review Letters}\ }\textbf {\bibinfo {volume} {125}},\ \bibinfo {pages} {268104} (\bibinfo {year} {2020})}\BibitemShut {NoStop}%
\bibitem [{\citenamefont {Chen}\ \emph {et~al.}(2024)\citenamefont {Chen}, \citenamefont {Lei}, \citenamefont {Xiang}, \citenamefont {Duan}, \citenamefont {Peng},\ and\ \citenamefont {Zhang}}]{chen2024emergent}%
  \BibitemOpen
  \bibfield  {author} {\bibinfo {author} {\bibfnamefont {J.}~\bibnamefont {Chen}}, \bibinfo {author} {\bibfnamefont {X.}~\bibnamefont {Lei}}, \bibinfo {author} {\bibfnamefont {Y.}~\bibnamefont {Xiang}}, \bibinfo {author} {\bibfnamefont {M.}~\bibnamefont {Duan}}, \bibinfo {author} {\bibfnamefont {X.}~\bibnamefont {Peng}},\ and\ \bibinfo {author} {\bibfnamefont {H.}~\bibnamefont {Zhang}},\ }\bibfield  {title} {\bibinfo {title} {Emergent chirality and hyperuniformity in an active mixture with nonreciprocal interactions},\ }\href@noop {} {\bibfield  {journal} {\bibinfo  {journal} {Physical Review Letters}\ }\textbf {\bibinfo {volume} {132}},\ \bibinfo {pages} {118301} (\bibinfo {year} {2024})}\BibitemShut {NoStop}%
\bibitem [{\citenamefont {Mart{\'\i}n-G{\'o}mez}\ \emph {et~al.}(2018)\citenamefont {Mart{\'\i}n-G{\'o}mez}, \citenamefont {Levis}, \citenamefont {D{\'\i}az-Guilera},\ and\ \citenamefont {Pagonabarraga}}]{martin2018collective}%
  \BibitemOpen
  \bibfield  {author} {\bibinfo {author} {\bibfnamefont {A.}~\bibnamefont {Mart{\'\i}n-G{\'o}mez}}, \bibinfo {author} {\bibfnamefont {D.}~\bibnamefont {Levis}}, \bibinfo {author} {\bibfnamefont {A.}~\bibnamefont {D{\'\i}az-Guilera}},\ and\ \bibinfo {author} {\bibfnamefont {I.}~\bibnamefont {Pagonabarraga}},\ }\bibfield  {title} {\bibinfo {title} {Collective motion of active brownian particles with polar alignment},\ }\href@noop {} {\bibfield  {journal} {\bibinfo  {journal} {Soft matter}\ }\textbf {\bibinfo {volume} {14}},\ \bibinfo {pages} {2610} (\bibinfo {year} {2018})}\BibitemShut {NoStop}%
\bibitem [{\citenamefont {Sese-Sansa}\ \emph {et~al.}(2018)\citenamefont {Sese-Sansa}, \citenamefont {Pagonabarraga},\ and\ \citenamefont {Levis}}]{sese2018velocity}%
  \BibitemOpen
  \bibfield  {author} {\bibinfo {author} {\bibfnamefont {E.}~\bibnamefont {Sese-Sansa}}, \bibinfo {author} {\bibfnamefont {I.}~\bibnamefont {Pagonabarraga}},\ and\ \bibinfo {author} {\bibfnamefont {D.}~\bibnamefont {Levis}},\ }\bibfield  {title} {\bibinfo {title} {Velocity alignment promotes motility-induced phase separation},\ }\href@noop {} {\bibfield  {journal} {\bibinfo  {journal} {Europhysics Letters}\ }\textbf {\bibinfo {volume} {124}},\ \bibinfo {pages} {30004} (\bibinfo {year} {2018})}\BibitemShut {NoStop}%
\bibitem [{\citenamefont {Kne{\v{z}}evi{\'c}}\ \emph {et~al.}(2022)\citenamefont {Kne{\v{z}}evi{\'c}}, \citenamefont {Welker},\ and\ \citenamefont {Stark}}]{knevzevic2022collective}%
  \BibitemOpen
  \bibfield  {author} {\bibinfo {author} {\bibfnamefont {M.}~\bibnamefont {Kne{\v{z}}evi{\'c}}}, \bibinfo {author} {\bibfnamefont {T.}~\bibnamefont {Welker}},\ and\ \bibinfo {author} {\bibfnamefont {H.}~\bibnamefont {Stark}},\ }\bibfield  {title} {\bibinfo {title} {Collective motion of active particles exhibiting non-reciprocal orientational interactions},\ }\href@noop {} {\bibfield  {journal} {\bibinfo  {journal} {Scientific Reports}\ }\textbf {\bibinfo {volume} {12}},\ \bibinfo {pages} {19437} (\bibinfo {year} {2022})}\BibitemShut {NoStop}%
\bibitem [{\citenamefont {Kumar}\ \emph {et~al.}(2024)\citenamefont {Kumar}, \citenamefont {Murali}, \citenamefont {Subramaniam}, \citenamefont {Singh},\ and\ \citenamefont {Thutupalli}}]{kumar2023emergent}%
  \BibitemOpen
  \bibfield  {author} {\bibinfo {author} {\bibfnamefont {M.}~\bibnamefont {Kumar}}, \bibinfo {author} {\bibfnamefont {A.}~\bibnamefont {Murali}}, \bibinfo {author} {\bibfnamefont {A.~G.}\ \bibnamefont {Subramaniam}}, \bibinfo {author} {\bibfnamefont {R.}~\bibnamefont {Singh}},\ and\ \bibinfo {author} {\bibfnamefont {S.}~\bibnamefont {Thutupalli}},\ }\bibfield  {title} {\bibinfo {title} {Emergent dynamics due to chemo-hydrodynamic self-interactions in active polymers},\ }\href {https://doi.org/10.1038/s41467-024-49155-7} {\bibfield  {journal} {\bibinfo  {journal} {Nature Commun.}\ }\textbf {\bibinfo {volume} {15}},\ \bibinfo {pages} {4903} (\bibinfo {year} {2024})}\BibitemShut {NoStop}%
\bibitem [{\citenamefont {Subramaniam}\ \emph {et~al.}(2024)\citenamefont {Subramaniam}, \citenamefont {Kumar}, \citenamefont {Thutupalli},\ and\ \citenamefont {Singh}}]{subramaniam2024rigid}%
  \BibitemOpen
  \bibfield  {author} {\bibinfo {author} {\bibfnamefont {A.~G.}\ \bibnamefont {Subramaniam}}, \bibinfo {author} {\bibfnamefont {M.}~\bibnamefont {Kumar}}, \bibinfo {author} {\bibfnamefont {S.}~\bibnamefont {Thutupalli}},\ and\ \bibinfo {author} {\bibfnamefont {R.}~\bibnamefont {Singh}},\ }\bibfield  {title} {\bibinfo {title} {Rigid flocks, undulatory gaits, and chiral foldamers in a chemically active polymer},\ }\href@noop {} {\bibfield  {journal} {\bibinfo  {journal} {New Journal of Physics}\ }\textbf {\bibinfo {volume} {26}},\ \bibinfo {pages} {083009} (\bibinfo {year} {2024})}\BibitemShut {NoStop}%
\bibitem [{\citenamefont {Bricard}\ \emph {et~al.}(2013)\citenamefont {Bricard}, \citenamefont {Caussin}, \citenamefont {Desreumaux}, \citenamefont {Dauchot},\ and\ \citenamefont {Bartolo}}]{bricard2013emergence}%
  \BibitemOpen
  \bibfield  {author} {\bibinfo {author} {\bibfnamefont {A.}~\bibnamefont {Bricard}}, \bibinfo {author} {\bibfnamefont {J.-B.}\ \bibnamefont {Caussin}}, \bibinfo {author} {\bibfnamefont {N.}~\bibnamefont {Desreumaux}}, \bibinfo {author} {\bibfnamefont {O.}~\bibnamefont {Dauchot}},\ and\ \bibinfo {author} {\bibfnamefont {D.}~\bibnamefont {Bartolo}},\ }\bibfield  {title} {\bibinfo {title} {Emergence of macroscopic directed motion in populations of motile colloids},\ }\href@noop {} {\bibfield  {journal} {\bibinfo  {journal} {Nature}\ }\textbf {\bibinfo {volume} {503}},\ \bibinfo {pages} {95} (\bibinfo {year} {2013})}\BibitemShut {NoStop}%
\bibitem [{\citenamefont {Das}\ \emph {et~al.}(2024)\citenamefont {Das}, \citenamefont {Ciarchi}, \citenamefont {Zhou}, \citenamefont {Yan}, \citenamefont {Zhang},\ and\ \citenamefont {Alert}}]{das2024flocking}%
  \BibitemOpen
  \bibfield  {author} {\bibinfo {author} {\bibfnamefont {S.}~\bibnamefont {Das}}, \bibinfo {author} {\bibfnamefont {M.}~\bibnamefont {Ciarchi}}, \bibinfo {author} {\bibfnamefont {Z.}~\bibnamefont {Zhou}}, \bibinfo {author} {\bibfnamefont {J.}~\bibnamefont {Yan}}, \bibinfo {author} {\bibfnamefont {J.}~\bibnamefont {Zhang}},\ and\ \bibinfo {author} {\bibfnamefont {R.}~\bibnamefont {Alert}},\ }\bibfield  {title} {\bibinfo {title} {Flocking by turning away},\ }\href@noop {} {\bibfield  {journal} {\bibinfo  {journal} {Physical Review X}\ }\textbf {\bibinfo {volume} {14}},\ \bibinfo {pages} {031008} (\bibinfo {year} {2024})}\BibitemShut {NoStop}%
\bibitem [{\citenamefont {Grossman}\ \emph {et~al.}(2008)\citenamefont {Grossman}, \citenamefont {Aranson},\ and\ \citenamefont {Jacob}}]{grossman2008emergence}%
  \BibitemOpen
  \bibfield  {author} {\bibinfo {author} {\bibfnamefont {D.}~\bibnamefont {Grossman}}, \bibinfo {author} {\bibfnamefont {I.}~\bibnamefont {Aranson}},\ and\ \bibinfo {author} {\bibfnamefont {E.~B.}\ \bibnamefont {Jacob}},\ }\bibfield  {title} {\bibinfo {title} {Emergence of agent swarm migration and vortex formation through inelastic collisions},\ }\href@noop {} {\bibfield  {journal} {\bibinfo  {journal} {New Journal of Physics}\ }\textbf {\bibinfo {volume} {10}},\ \bibinfo {pages} {023036} (\bibinfo {year} {2008})}\BibitemShut {NoStop}%
\bibitem [{\citenamefont {Hanke}\ \emph {et~al.}(2013)\citenamefont {Hanke}, \citenamefont {Weber},\ and\ \citenamefont {Frey}}]{hanke2013understanding}%
  \BibitemOpen
  \bibfield  {author} {\bibinfo {author} {\bibfnamefont {T.}~\bibnamefont {Hanke}}, \bibinfo {author} {\bibfnamefont {C.~A.}\ \bibnamefont {Weber}},\ and\ \bibinfo {author} {\bibfnamefont {E.}~\bibnamefont {Frey}},\ }\bibfield  {title} {\bibinfo {title} {Understanding collective dynamics of soft active colloids by binary scattering},\ }\href@noop {} {\bibfield  {journal} {\bibinfo  {journal} {Physical Review E—Statistical, Nonlinear, and Soft Matter Physics}\ }\textbf {\bibinfo {volume} {88}},\ \bibinfo {pages} {052309} (\bibinfo {year} {2013})}\BibitemShut {NoStop}%
\bibitem [{\citenamefont {Lam}\ \emph {et~al.}(2015)\citenamefont {Lam}, \citenamefont {Schindler},\ and\ \citenamefont {Dauchot}}]{lam2015self}%
  \BibitemOpen
  \bibfield  {author} {\bibinfo {author} {\bibfnamefont {K.-D. N.~T.}\ \bibnamefont {Lam}}, \bibinfo {author} {\bibfnamefont {M.}~\bibnamefont {Schindler}},\ and\ \bibinfo {author} {\bibfnamefont {O.}~\bibnamefont {Dauchot}},\ }\bibfield  {title} {\bibinfo {title} {Self-propelled hard disks: implicit alignment and transition to collective motion},\ }\href@noop {} {\bibfield  {journal} {\bibinfo  {journal} {New Journal of Physics}\ }\textbf {\bibinfo {volume} {17}},\ \bibinfo {pages} {113056} (\bibinfo {year} {2015})}\BibitemShut {NoStop}%
\bibitem [{\citenamefont {Miranda}\ \emph {et~al.}(2025)\citenamefont {Miranda}, \citenamefont {Levis},\ and\ \citenamefont {Valeriani}}]{miranda2025collective}%
  \BibitemOpen
  \bibfield  {author} {\bibinfo {author} {\bibfnamefont {J.~P.}\ \bibnamefont {Miranda}}, \bibinfo {author} {\bibfnamefont {D.}~\bibnamefont {Levis}},\ and\ \bibinfo {author} {\bibfnamefont {C.}~\bibnamefont {Valeriani}},\ }\bibfield  {title} {\bibinfo {title} {Collective motion of energy depot active disks},\ }\href@noop {} {\bibfield  {journal} {\bibinfo  {journal} {Soft Matter}\ }\textbf {\bibinfo {volume} {21}},\ \bibinfo {pages} {1045} (\bibinfo {year} {2025})}\BibitemShut {NoStop}%
\bibitem [{\citenamefont {Caprini}\ and\ \citenamefont {L{\"o}wen}(2023)}]{caprini2023flocking}%
  \BibitemOpen
  \bibfield  {author} {\bibinfo {author} {\bibfnamefont {L.}~\bibnamefont {Caprini}}\ and\ \bibinfo {author} {\bibfnamefont {H.}~\bibnamefont {L{\"o}wen}},\ }\bibfield  {title} {\bibinfo {title} {Flocking without alignment interactions in attractive active brownian particles},\ }\href@noop {} {\bibfield  {journal} {\bibinfo  {journal} {Physical Review Letters}\ }\textbf {\bibinfo {volume} {130}},\ \bibinfo {pages} {148202} (\bibinfo {year} {2023})}\BibitemShut {NoStop}%
\bibitem [{\citenamefont {Camley}\ and\ \citenamefont {Rappel}(2017)}]{camley2017physical}%
  \BibitemOpen
  \bibfield  {author} {\bibinfo {author} {\bibfnamefont {B.~A.}\ \bibnamefont {Camley}}\ and\ \bibinfo {author} {\bibfnamefont {W.-J.}\ \bibnamefont {Rappel}},\ }\bibfield  {title} {\bibinfo {title} {Physical models of collective cell motility: from cell to tissue},\ }\href@noop {} {\bibfield  {journal} {\bibinfo  {journal} {Journal of physics D: Applied physics}\ }\textbf {\bibinfo {volume} {50}},\ \bibinfo {pages} {113002} (\bibinfo {year} {2017})}\BibitemShut {NoStop}%
\bibitem [{\citenamefont {Hayakawa}\ \emph {et~al.}(2020)\citenamefont {Hayakawa}, \citenamefont {Hiraiwa}, \citenamefont {Wada}, \citenamefont {Kuwayama},\ and\ \citenamefont {Shibata}}]{hayakawa2020polar}%
  \BibitemOpen
  \bibfield  {author} {\bibinfo {author} {\bibfnamefont {M.}~\bibnamefont {Hayakawa}}, \bibinfo {author} {\bibfnamefont {T.}~\bibnamefont {Hiraiwa}}, \bibinfo {author} {\bibfnamefont {Y.}~\bibnamefont {Wada}}, \bibinfo {author} {\bibfnamefont {H.}~\bibnamefont {Kuwayama}},\ and\ \bibinfo {author} {\bibfnamefont {T.}~\bibnamefont {Shibata}},\ }\bibfield  {title} {\bibinfo {title} {Polar pattern formation induced by contact following locomotion in a multicellular system},\ }\href@noop {} {\bibfield  {journal} {\bibinfo  {journal} {Elife}\ }\textbf {\bibinfo {volume} {9}},\ \bibinfo {pages} {e53609} (\bibinfo {year} {2020})}\BibitemShut {NoStop}%
\bibitem [{\citenamefont {Pohl}\ and\ \citenamefont {Stark}(2014)}]{pohlStarkPRL2014}%
  \BibitemOpen
  \bibfield  {author} {\bibinfo {author} {\bibfnamefont {O.}~\bibnamefont {Pohl}}\ and\ \bibinfo {author} {\bibfnamefont {H.}~\bibnamefont {Stark}},\ }\bibfield  {title} {\bibinfo {title} {Dynamic clustering and chemotactic collapse of self-phoretic active particles},\ }\href {https://doi.org/10.1103/PhysRevLett.112.238303} {\bibfield  {journal} {\bibinfo  {journal} {Phys. Rev. Lett.}\ }\textbf {\bibinfo {volume} {112}},\ \bibinfo {pages} {238303} (\bibinfo {year} {2014})}\BibitemShut {NoStop}%
\bibitem [{\citenamefont {Hokmabad}\ \emph {et~al.}(2022)\citenamefont {Hokmabad}, \citenamefont {Agudo-Canalejo}, \citenamefont {Saha}, \citenamefont {Golestanian},\ and\ \citenamefont {Maass}}]{hokmabad2022chemotactic}%
  \BibitemOpen
  \bibfield  {author} {\bibinfo {author} {\bibfnamefont {B.~V.}\ \bibnamefont {Hokmabad}}, \bibinfo {author} {\bibfnamefont {J.}~\bibnamefont {Agudo-Canalejo}}, \bibinfo {author} {\bibfnamefont {S.}~\bibnamefont {Saha}}, \bibinfo {author} {\bibfnamefont {R.}~\bibnamefont {Golestanian}},\ and\ \bibinfo {author} {\bibfnamefont {C.~C.}\ \bibnamefont {Maass}},\ }\bibfield  {title} {\bibinfo {title} {Chemotactic self-caging in active emulsions},\ }\href@noop {} {\bibfield  {journal} {\bibinfo  {journal} {Proc. Natl. Acad. Sci.}\ }\textbf {\bibinfo {volume} {119}},\ \bibinfo {pages} {e2122269119} (\bibinfo {year} {2022})}\BibitemShut {NoStop}%
\bibitem [{\citenamefont {Dwivedi}\ \emph {et~al.}(2022)\citenamefont {Dwivedi}, \citenamefont {Pillai},\ and\ \citenamefont {Mangal}}]{dwivedi2022self}%
  \BibitemOpen
  \bibfield  {author} {\bibinfo {author} {\bibfnamefont {P.}~\bibnamefont {Dwivedi}}, \bibinfo {author} {\bibfnamefont {D.}~\bibnamefont {Pillai}},\ and\ \bibinfo {author} {\bibfnamefont {R.}~\bibnamefont {Mangal}},\ }\bibfield  {title} {\bibinfo {title} {Self-propelled swimming droplets},\ }\href@noop {} {\bibfield  {journal} {\bibinfo  {journal} {Current Opinion in Colloid \& Interface Science}\ }\textbf {\bibinfo {volume} {61}},\ \bibinfo {pages} {101614} (\bibinfo {year} {2022})}\BibitemShut {NoStop}%
\bibitem [{\citenamefont {Liebchen}\ and\ \citenamefont {L{\"o}wen}(2019)}]{liebchen2019interactions}%
  \BibitemOpen
  \bibfield  {author} {\bibinfo {author} {\bibfnamefont {B.}~\bibnamefont {Liebchen}}\ and\ \bibinfo {author} {\bibfnamefont {H.}~\bibnamefont {L{\"o}wen}},\ }\bibfield  {title} {\bibinfo {title} {Which interactions dominate in active colloids?},\ }\href@noop {} {\bibfield  {journal} {\bibinfo  {journal} {The Journal of chemical physics}\ }\textbf {\bibinfo {volume} {150}} (\bibinfo {year} {2019})}\BibitemShut {NoStop}%
\bibitem [{\citenamefont {Saha}\ \emph {et~al.}(2014)\citenamefont {Saha}, \citenamefont {Golestanian},\ and\ \citenamefont {Ramaswamy}}]{saha2014clusters}%
  \BibitemOpen
  \bibfield  {author} {\bibinfo {author} {\bibfnamefont {S.}~\bibnamefont {Saha}}, \bibinfo {author} {\bibfnamefont {R.}~\bibnamefont {Golestanian}},\ and\ \bibinfo {author} {\bibfnamefont {S.}~\bibnamefont {Ramaswamy}},\ }\bibfield  {title} {\bibinfo {title} {Clusters, asters, and collective oscillations in chemotactic colloids},\ }\href@noop {} {\bibfield  {journal} {\bibinfo  {journal} {Phys. Rev. E}\ }\textbf {\bibinfo {volume} {89}},\ \bibinfo {pages} {062316} (\bibinfo {year} {2014})}\BibitemShut {NoStop}%
\bibitem [{\citenamefont {Soto}\ and\ \citenamefont {Golestanian}(2015)}]{soto2015self}%
  \BibitemOpen
  \bibfield  {author} {\bibinfo {author} {\bibfnamefont {R.}~\bibnamefont {Soto}}\ and\ \bibinfo {author} {\bibfnamefont {R.}~\bibnamefont {Golestanian}},\ }\bibfield  {title} {\bibinfo {title} {Self-assembly of active colloidal molecules with dynamic function},\ }\href@noop {} {\bibfield  {journal} {\bibinfo  {journal} {Phys. Rev. E}\ }\textbf {\bibinfo {volume} {91}},\ \bibinfo {pages} {052304} (\bibinfo {year} {2015})}\BibitemShut {NoStop}%
\bibitem [{\citenamefont {R{\"u}ckner}\ and\ \citenamefont {Kapral}(2007)}]{ruckner2007chemically}%
  \BibitemOpen
  \bibfield  {author} {\bibinfo {author} {\bibfnamefont {G.}~\bibnamefont {R{\"u}ckner}}\ and\ \bibinfo {author} {\bibfnamefont {R.}~\bibnamefont {Kapral}},\ }\bibfield  {title} {\bibinfo {title} {Chemically powered nanodimers},\ }\href@noop {} {\bibfield  {journal} {\bibinfo  {journal} {Physical review letters}\ }\textbf {\bibinfo {volume} {98}},\ \bibinfo {pages} {150603} (\bibinfo {year} {2007})}\BibitemShut {NoStop}%
\bibitem [{\citenamefont {Singh}\ \emph {et~al.}(2019)\citenamefont {Singh}, \citenamefont {Adhikari},\ and\ \citenamefont {Cates}}]{singh2019competing}%
  \BibitemOpen
  \bibfield  {author} {\bibinfo {author} {\bibfnamefont {R.}~\bibnamefont {Singh}}, \bibinfo {author} {\bibfnamefont {R.}~\bibnamefont {Adhikari}},\ and\ \bibinfo {author} {\bibfnamefont {M.}~\bibnamefont {Cates}},\ }\bibfield  {title} {\bibinfo {title} {Competing chemical and hydrodynamic interactions in autophoretic colloidal suspensions},\ }\href@noop {} {\bibfield  {journal} {\bibinfo  {journal} {The Journal of chemical physics}\ }\textbf {\bibinfo {volume} {151}} (\bibinfo {year} {2019})}\BibitemShut {NoStop}%
\bibitem [{\citenamefont {Turk}\ \emph {et~al.}(2024)\citenamefont {Turk}, \citenamefont {Adhikari},\ and\ \citenamefont {Singh}}]{turk2024fluctuating}%
  \BibitemOpen
  \bibfield  {author} {\bibinfo {author} {\bibfnamefont {G.}~\bibnamefont {Turk}}, \bibinfo {author} {\bibfnamefont {R.}~\bibnamefont {Adhikari}},\ and\ \bibinfo {author} {\bibfnamefont {R.}~\bibnamefont {Singh}},\ }\bibfield  {title} {\bibinfo {title} {Fluctuating hydrodynamics of an autophoretic particle near a permeable interface},\ }\href@noop {} {\bibfield  {journal} {\bibinfo  {journal} {Journal of Fluid Mechanics}\ }\textbf {\bibinfo {volume} {998}},\ \bibinfo {pages} {A34} (\bibinfo {year} {2024})}\BibitemShut {NoStop}%
\bibitem [{\citenamefont {Geyer}\ \emph {et~al.}(2018)\citenamefont {Geyer}, \citenamefont {Morin},\ and\ \citenamefont {Bartolo}}]{geyer2018sounds}%
  \BibitemOpen
  \bibfield  {author} {\bibinfo {author} {\bibfnamefont {D.}~\bibnamefont {Geyer}}, \bibinfo {author} {\bibfnamefont {A.}~\bibnamefont {Morin}},\ and\ \bibinfo {author} {\bibfnamefont {D.}~\bibnamefont {Bartolo}},\ }\bibfield  {title} {\bibinfo {title} {Sounds and hydrodynamics of polar active fluids},\ }\href@noop {} {\bibfield  {journal} {\bibinfo  {journal} {Nature materials}\ }\textbf {\bibinfo {volume} {17}},\ \bibinfo {pages} {789} (\bibinfo {year} {2018})}\BibitemShut {NoStop}%
\bibitem [{\citenamefont {Geyer}\ \emph {et~al.}(2019)\citenamefont {Geyer}, \citenamefont {Martin}, \citenamefont {Tailleur},\ and\ \citenamefont {Bartolo}}]{geyer2019freezing}%
  \BibitemOpen
  \bibfield  {author} {\bibinfo {author} {\bibfnamefont {D.}~\bibnamefont {Geyer}}, \bibinfo {author} {\bibfnamefont {D.}~\bibnamefont {Martin}}, \bibinfo {author} {\bibfnamefont {J.}~\bibnamefont {Tailleur}},\ and\ \bibinfo {author} {\bibfnamefont {D.}~\bibnamefont {Bartolo}},\ }\bibfield  {title} {\bibinfo {title} {Freezing a flock: Motility-induced phase separation in polar active liquids},\ }\href@noop {} {\bibfield  {journal} {\bibinfo  {journal} {Physical Review X}\ }\textbf {\bibinfo {volume} {9}},\ \bibinfo {pages} {031043} (\bibinfo {year} {2019})}\BibitemShut {NoStop}%
\bibitem [{siT()}]{siText}%
  \BibitemOpen
  \href@noop {} {\bibinfo {title} {See the supplemental material at this {URL}: [to be inserted].}}\BibitemShut {Stop}%
\bibitem [{\citenamefont {Zhang}\ \emph {et~al.}(2021)\citenamefont {Zhang}, \citenamefont {Alert}, \citenamefont {Yan}, \citenamefont {Wingreen},\ and\ \citenamefont {Granick}}]{zhang2021active}%
  \BibitemOpen
  \bibfield  {author} {\bibinfo {author} {\bibfnamefont {J.}~\bibnamefont {Zhang}}, \bibinfo {author} {\bibfnamefont {R.}~\bibnamefont {Alert}}, \bibinfo {author} {\bibfnamefont {J.}~\bibnamefont {Yan}}, \bibinfo {author} {\bibfnamefont {N.~S.}\ \bibnamefont {Wingreen}},\ and\ \bibinfo {author} {\bibfnamefont {S.}~\bibnamefont {Granick}},\ }\bibfield  {title} {\bibinfo {title} {Active phase separation by turning towards regions of higher density},\ }\href@noop {} {\bibfield  {journal} {\bibinfo  {journal} {Nature Physics}\ }\textbf {\bibinfo {volume} {17}},\ \bibinfo {pages} {961} (\bibinfo {year} {2021})}\BibitemShut {NoStop}%
\bibitem [{\citenamefont {Chaikin}\ \emph {et~al.}(1995)\citenamefont {Chaikin}, \citenamefont {Lubensky},\ and\ \citenamefont {Witten}}]{chaikin1995principles}%
  \BibitemOpen
  \bibfield  {author} {\bibinfo {author} {\bibfnamefont {P.~M.}\ \bibnamefont {Chaikin}}, \bibinfo {author} {\bibfnamefont {T.~C.}\ \bibnamefont {Lubensky}},\ and\ \bibinfo {author} {\bibfnamefont {T.~A.}\ \bibnamefont {Witten}},\ }\href@noop {} {\emph {\bibinfo {title} {Principles of condensed matter physics}}},\ Vol.~\bibinfo {volume} {10}\ (\bibinfo  {publisher} {Cambridge university press Cambridge},\ \bibinfo {year} {1995})\BibitemShut {NoStop}%
\bibitem [{\citenamefont {K{\"u}rsten}\ and\ \citenamefont {Ihle}(2020)}]{kursten2020dry}%
  \BibitemOpen
  \bibfield  {author} {\bibinfo {author} {\bibfnamefont {R.}~\bibnamefont {K{\"u}rsten}}\ and\ \bibinfo {author} {\bibfnamefont {T.}~\bibnamefont {Ihle}},\ }\bibfield  {title} {\bibinfo {title} {Dry active matter exhibits a self-organized cross sea phase},\ }\href@noop {} {\bibfield  {journal} {\bibinfo  {journal} {Physical Review Letters}\ }\textbf {\bibinfo {volume} {125}},\ \bibinfo {pages} {188003} (\bibinfo {year} {2020})}\BibitemShut {NoStop}%
\bibitem [{\citenamefont {Wensink}\ and\ \citenamefont {L{\"o}wen}(2012)}]{wensink2012emergent}%
  \BibitemOpen
  \bibfield  {author} {\bibinfo {author} {\bibfnamefont {H.}~\bibnamefont {Wensink}}\ and\ \bibinfo {author} {\bibfnamefont {H.}~\bibnamefont {L{\"o}wen}},\ }\bibfield  {title} {\bibinfo {title} {Emergent states in dense systems of active rods: from swarming to turbulence},\ }\href@noop {} {\bibfield  {journal} {\bibinfo  {journal} {Journal of Physics: Condensed Matter}\ }\textbf {\bibinfo {volume} {24}},\ \bibinfo {pages} {464130} (\bibinfo {year} {2012})}\BibitemShut {NoStop}%
\bibitem [{\citenamefont {Aditi~Simha}\ and\ \citenamefont {Ramaswamy}(2002)}]{aditi2002hydrodynamic}%
  \BibitemOpen
  \bibfield  {author} {\bibinfo {author} {\bibfnamefont {R.}~\bibnamefont {Aditi~Simha}}\ and\ \bibinfo {author} {\bibfnamefont {S.}~\bibnamefont {Ramaswamy}},\ }\bibfield  {title} {\bibinfo {title} {Hydrodynamic fluctuations and instabilities in ordered suspensions of self-propelled particles},\ }\href@noop {} {\bibfield  {journal} {\bibinfo  {journal} {Physical review letters}\ }\textbf {\bibinfo {volume} {89}},\ \bibinfo {pages} {058101} (\bibinfo {year} {2002})}\BibitemShut {NoStop}%
\bibitem [{\citenamefont {Thutupalli}\ \emph {et~al.}(2018)\citenamefont {Thutupalli}, \citenamefont {Geyer}, \citenamefont {Singh}, \citenamefont {Adhikari},\ and\ \citenamefont {Stone}}]{thutupalli2018flow}%
  \BibitemOpen
  \bibfield  {author} {\bibinfo {author} {\bibfnamefont {S.}~\bibnamefont {Thutupalli}}, \bibinfo {author} {\bibfnamefont {D.}~\bibnamefont {Geyer}}, \bibinfo {author} {\bibfnamefont {R.}~\bibnamefont {Singh}}, \bibinfo {author} {\bibfnamefont {R.}~\bibnamefont {Adhikari}},\ and\ \bibinfo {author} {\bibfnamefont {H.~A.}\ \bibnamefont {Stone}},\ }\bibfield  {title} {\bibinfo {title} {Flow-induced phase separation of active particles is controlled by boundary conditions},\ }\href@noop {} {\bibfield  {journal} {\bibinfo  {journal} {Proc. Natl. Acad. Sci.}\ }\textbf {\bibinfo {volume} {115}},\ \bibinfo {pages} {5403} (\bibinfo {year} {2018})}\BibitemShut {NoStop}%
\end{thebibliography}
%

\end{document}